\documentclass[12pt]{article}
\usepackage{epsfig}
\usepackage{cite}
\usepackage{float}
\usepackage{mathrsfs}
\usepackage{amsfonts}
\usepackage{array}
\usepackage{epsfig}
\usepackage{amsmath}    
\usepackage{amssymb}   
\usepackage{graphicx}   
\usepackage{verbatim}   
\usepackage{color}      
\usepackage{subfigure}  
\usepackage{subfiles}
\usepackage{slashed}

\def\beq{\begin{equation}}
\def\eeq{\end{equation}}
\def\beqa{\begin{eqnarray}}
\def\eeqa{\end{eqnarray}}
\def\ba{\begin{eqnarray}}
\def\ea{\end{eqnarray}}
 
\newlength{\dinwidth} \newlength{\dinmargin}
\begin{document}

\begin{center}
{\Large \bf Precision studies \\\medskip for string derived $Z'$ dynamics at the LHC}
\end{center}
\vspace{2mm}
\begin{center}
{\large Andrew McEntaggart$^{(a,b)}$, Alon E. Faraggi$^{(c)}$, Marco Guzzi$^{(a)}$}\\
\vspace{2mm}
{\it $^{(a)}$Department of Physics, Kennesaw State University,\\
Kennesaw, GA 30144, USA}

\it{$^{(b)}$Department of Physics, Georgia Institute of Technology, \\Atlanta, GA 30332, USA}

\it{$^{(c)}$Department of Mathematical Sciences,
University of Liverpool,\\
Liverpool L69 7ZL, UK}
\end{center}

\begin{abstract}
We consider $Z'$s in heterotic string derived models and
study $Z'$ resonant production at the TeV scale at the Large Hadron Collider (LHC).
We use various kinematic differential distributions for the Drell-Yan process
at NNLO in QCD to explore the parameter space of such models and investigate $Z'$ couplings. 
In particular, we study the impact of $Z$-$Z'$ kinetic-mixing
interactions on forward-backward asymmetry ($A_{FB}$) and other distributions at the LHC.
\end{abstract}

\section{Introduction}

With the observation of a scalar particle compatible with the Standard Model electroweak Higgs doublet,
the last piece in the Standard Model has been confirmed experimentally, and in the coming years
its basic properties will be probed and elucidated. Furthermore, the discovery of the Higgs 
particle at 125 GeV suggests that the electroweak symmetry breaking mechanism is perturbative,
rather than non-perturbative. This supports the old expectation that the Standard Model may
provide viable parameterisation of particle physics data up to the Planck scale, where 
gravitational effects break the effective field theory description. Alas, the 
question remains how the Higgs mass and the ensuing electroweak symmetry breaking scales are
protected from radiative corrections from the higher energy scale. While not yet 
seen at low energy experiments, supersymmetry remains one of the most appealing extensions
of the Standard Model to address this puzzle. Furthermore, many supersymmetric models
in fact predict that the Higgs particle should exist below 135 GeV. The observed Higgs particle 
is therefore compatible with supersymmetry. However, while supersymmetry
protects the Higgs mass at the multi--loop level in the perturbative expansion, it introduces
a problem at tree level, which is known as the $\mu$--problem. Supersymmetry mandates the existence
of a pair of Higgs doublets, which can receive a bilinear mass term in the superpotential. In 
the Minimal Supersymmetric Standard Model, the $\mu$--parameter is therefore set by hand
to be of the order of the electroweak scale, but nothing prevents it from being of the 
order of the Planck scale. Furthermore, absence of global symmetries in quantum gravity and 
the experience from quasi--realistic string derived models, do in fact suggest that generically
the $\mu$--parameter will be of that order. A possible solution to the problem may be obtained 
if the Higgs states are chiral under an additional $U(1)$ symmetry, which remains unbroken 
down to TeV scale. In this case, the bi--linear mass term for the supersymmetric Higgs pair
can only be generated by the vacuum Expectation Value (VEV) that breaks the extra $U(1)$ symmetry. 

String theory provides the most developed framework to study the incorporation of the 
Standard Model in a perturbatively consistent theory of quantum gravity. Indeed, over the
past few decades detailed phenomenological models have been built that produce the spectrum
of the Minimal Supersymmetric Standard Model (MSSM) in the effective low energy field theory
limit below the Planck scale (for review and references see {\it e.g.} \cite{Ibanez:2012zz}). 
The heterotic--string models constructed in the 
free fermionic formulation 
\cite{Faraggi:1989ka, Cleaver:1998saa, Faraggi:1992fa, Faraggi:2017cnh}, 
which are $Z_2\times Z_2$ orbifolds of six dimensional 
toroidal spaces \cite{Athanasopoulos:2016aws}, are among the most realistic models constructed to date, and give 
rise to an abundance of three generation models with different unbroken $SO(10)$ subgroups 
below the string scale. However, the construction of string models with 
additional $U(1)$ gauge symmetries, that may stay
unbroken down to low scales, has proven to be a difficult
task \cite{Faraggi:2011xu, Faraggi:2013nia}. The reason is that the extra $U(1)$ gauge symmetries
that are often discussed in the context of Grand Unified Theory (GUT) 
extensions of the Standard Model are anomalous in explicit string
models and therefore cannot remain unbroken down to
low scales. In fact to date, there exist a single string derived model 
in which the extra $U(1)$ is anomaly free and can remain unbroken
down to the TeV scale \cite{Faraggi:2014ica}. 
As a bonus the string derived model predicts the existence
of electroweak doublet pairs at the $Z^\prime$ breaking scale, that are 
chiral under the extra $U(1)$ symmetry. The bilinear mass term for these 
electroweak Higgs doublets can therefore be generated only by the VEV that 
breaks the extra $U(1)$ symmetry. 

In this paper, we therefore study some of the properties of the extra $U(1)$
symmetry predicted by the string model and explore the parameter space of the $Z'$ 
by using various kinematic distributions of resonant lepton-pair production in the Drell-Yan (DY) process at the LHC. Other related literature can be found in refs.~\cite{Komachenko:1989qn,Das:2021esm,Anastasopoulos:2008jt}. In addition, we study the effect of gauge kinetic-mixing interactions $\chi F^{\mu\nu}F'_{\mu\nu}$ in the lagrangian~\cite{Holdom:1985ag}.
The impact of such a renormalizable operator is interesting in this context because it may in general be produced (at both field theory or string theory level) by physics at high scales above the $Z'$ breaking scale, with no suppression by the large-mass scale~\cite{Polchinski:1982an,Dienes:1996zr} and can impact the couplings of $Z'$s at the TeV scale~\cite{Rizzo:1998ut}. The interplay between the gauge kinetic-mixing parameter and coupling of the $Z'$, as well as other $Z'$ properties, can be explored by using forward-backward asymmetry ($A_{FB}$) distributions in DY, where this observable can be advantageous in extra-resonance searches and as a model discrimination tool~\cite{London:1986dk,Rosner:1986cv,Hewett:1988xc,Cvetic:1995zs,Rosner:1995ft,Dittmar:1996my,Bodek:2001xk,CDF:1997wdd,Accomando:2015cfa,Fiaschi:2022wgl,Ball:2022qtp}.

This paper is organized as follows. In Sec.~\ref{string-model} we describe the structure of the string derived $Z^\prime$ model. In Sec.~\ref{pheno} we discuss the phenomenological aspects of the model and analyze kinetic mixing. In Sec.~\ref{numerical} we illustrate and discuss our results. Finally, in Sec.~\ref{conclusions} we present our conclusions.

\section{The String Model~\label{string-model}}

We elaborate in this section on the structure of the string derived $Z^\prime$ model
\cite{Faraggi:2014ica}. 
The model was constructed in the free fermionic
formulation~\cite{Antoniadis:1986rn,Kawai:1986ah,Antoniadis:1987wp}.
Only the important points relevant for the properties 
of the extra $Z^\prime$ gauge boson at low scales are discussed here 
and further details can be found in the original literature. 
Particularly relevant for the low energy phenomenology is
the precise combination of worldsheet $U(1)$ currents 
predicted by the string model. In this respect we note that while 
similar aspects can be discussed in field theory models, in the 
string model the $U(1)$ combination is forced due to the available 
scalar multiplets to break the GUT symmetry, whereas in field theory
models the scalar multiplets are a matter of choice. 

In the free fermionic formulation 
all the worldsheet degrees of freedom required to cancel the conformal anomaly
are represented in terms of free fermions propogating
on the string worldsheet. In the standard notation the 64 worldsheet
fermions in the lightcone gauge are denoted as:
\leftline{${\underline{{\hbox{Left-Movers}}}}$:~~$\psi^\mu,~~{ \chi_i},
~~{ y_i,~~\omega_i}~~~~(\mu=1,2,~i=1,\cdots,6)$}
\vspace{4mm}
\leftline{${\underline{{\hbox{Right-Movers}}}}$}
$${\bar\phi}_{A=1,\cdots,44}=
\begin{cases}
~~{ {\bar y}_i~,~ {\bar\omega}_i} & i=1,{\cdots},6\cr
  & \cr
~~{ {\bar\eta}_i} & i=1,2,3~~\cr
~~{ {\bar\psi}_{1,\cdots,5}} & \cr
~~{{\bar\phi}_{1,\cdots,8}}  &
\end{cases}
$$
where the $\{y,\omega\vert{\bar y},{\bar\omega}\}^{1,\cdots,6}$
correspond to the
six dimensions of the internally compactified manifold;
${\bar\psi}^{1,\cdots,5}$ generate the $SO(10)$ GUT symmetry;
${\bar\phi}^{1,\cdots,8}$ generate the
hidden sector gauge group; and ${\bar\eta}^{1,2,3}$
generate three $U(1)$ gauge symmetries.
Models in the free fermionic formulation are defined in terms of boundary
condition basis vectors, which specify the transformation properties
of the worldsheet fermions around the noncontractible loops of the vacuum 
to vacuum amplitude, and the Generalised GSO projection coefficients of the one loop
partition function
\cite{Antoniadis:1986rn, Kawai:1986ah, Antoniadis:1987wp}.
The free fermion models are
$Z_2\times Z_2$ orbifold of six dimensional toroidal 
manifolds with discrete Wilson lines
\cite{Athanasopoulos:2016aws}.

The Standard Model particle charges and data motivates the embedding
of the Standard Model states in representations of Grand Unified Theories, 
like $SO(10)$ and $E_6$. The rank of these groups exceeds that of 
the Standard Model, and therefore they predict the existence of 
additional gauge symmetries, beyond the Standard Model. 
Compactifications of the heterotic--string to four 
dimensions do in fact give rise to $E_6$ GUT like models \cite{Candelas:1985en}, 
which gives rise to string inspired $Z^\prime$ models.
The additional $U(1)$ symmetries
in these string inspired models have an $E_6$ embedding
and produced numerous papers since the mid--eighties (for reviews and references see {\it e.g.}~\cite{Zwirner:1987kxa,King:2020ldn,Hewett:1988xc,Leike:1998wr,Komachenko:1989qn}).
The construction of string derived models that allow for an unbroken
extra $U(1)$ symmetry to remain unbroken down to low scales proves, however,
to be very difficult.
The symmetry breaking pattern $E_6\rightarrow SO(10)\times U(1)_A$ 
in the string derived 
models entails that $U(1)_A$
is anomalous and cannot be part of an unbroken $U(1)_{Z^\prime}$ at low
energy scales \cite{Cleaver:1997rk}.
String derived models with anomaly free 
$U(1)_{Z^\prime}\notin E_6$ were analysed in~\cite{Pati:1996fn,Faraggi:2000cm,Coriano:2007ba,Faraggi:2011xu},
but agreement with the measured values of $\sin^2(\theta)_W(M_Z)$ and
$\alpha_s(M_Z)$ favours $Z^\prime$ models with $E_6$ embedding~\cite{Faraggi:2013nia}.
We remark that the anomaly free $U(1)$ combination of
$U(1)_{B-L}$ and $U(1)_{T_{3_R}}\in SO(10)$ may in principle
stay unbroken down to low energy scales \cite{Faraggi:1990ita}.
However, adequate suppression of the left--handed neutrino masses
is facilitated if this $U(1)$ symmetry is broken at a high energy
scale \cite{Faraggi:1990it}. Constructing string models that enable
an extra $U(1)\in E_6$ symmetry to stay unbroken down to low energy scales
necessitates the construction of string models in which
$U(1)_A$ is anomaly free. One method of achieving this outcome
is to enhance $U(1)_A$ to a non-Abelian gauge symmetry {\'a} la
ref. \cite{Bernard:2012vf}.
An alternative is the string derived model of ref. \cite{Faraggi:2014ica}
that uses the Spinor--Vector Duality (SVD) that was observed in
$Z_2\times Z_2$ orbifolds
\cite{Faraggi:2006pk,Angelantonj:2010zj, Faraggi:2011aw}.
The SVD is under the
exchange of the total number of $(16+\overline{16})$ representations
of $SO(10)$ with the total number of $10$ representations, and
is readily understood if we consider the enhancement of
$SO(10)\times U(1)_A$ to $E_6$.
The chiral and anti--chiral multiplets
of $E_6$ decompose under $SO(10)\times U(1)$ as $27=16+10+1$ and
$\overline{27}= \overline{16}+10+1$.
In this case the $\#_1$ of $(16+\overline{16})$ and $\#_2$ of $10$
multiplets are equal.
The $E_6$ symmetry point in the moduli space
corresponds to a self--dual point under the exchange
of the total number of $SO(10)$ spinorial plus anti--spinorial, with the
total number of vectorial, multiplets.
Breaking the $E_6$ gauge symmetry to $SO(10)\times U(1)_A$ results in the projection
of some of the spinorial and vectorial multiplets, which results in $U(1)_A$ being anomalous.
However, there may exist vacua with equal numbers of
$(16+\overline{16})$ spinorial, and $10$ vectorial, multiplets,
and traceless $U(1)_A$, without enhancement of the $SO(10)\times U(1)_A$
symmetry to $E_6$ \cite{Faraggi:2014ica}.
In such models the chiral spectrum still forms
complete $E_6$ representations, but the gauge symmetry is not
enhanced to $E_6$. In this cases
$U(1)_A$ is anomaly free and may remain unbroken down to low
scales.

A fishing algorithm to extract models
with specified physical properties was developed by using the free fermionic 
model building rules
\cite{Faraggi:2004rq,Assel:2010wj,Faraggi:2014hqa,Faraggi:2017cnh,Faraggi:2018hqx, Faraggi:2019qoq,Faraggi:2020wej,Faraggi:2020wld}.
In ref. \cite{Faraggi:2014ica}, using the free fermionic fishing algorithm, such
a spinor--vector self dual model was obtained with subsequent breaking at the string scale
of the $SO(10)$ symmetry to the $SO(6)\times SO(4)$ subgroup, and preserves the
spinor--vector self--duality. This model is a string derived model
in which an extra $U(1)$ with $E_6$ embedding may remain unbroken down to
low scales. The full massless spectrum of the string derived model is given
in ref. \cite{Faraggi:2014ica}.
The observable and hidden gauge groups at the
string scale are produced by untwisted sector states and are given by:
\begin{eqnarray}
{\rm observable} ~: &~~SO(6)\times SO(4) \times
U(1)_1 \times U(1)_2\times U(1)_3 \nonumber\\
{\rm hidden}     ~: &SO(4)^2\times SO(8)~~~~~~~~~~~~~~~~~~~~~~~~~~~~~~~\nonumber
\end{eqnarray}
The massless string spectrum contains the
fields required to break the GUT symmetry to the Standard Model.


The string model contains two anomalous $U(1)$ symmetries
\begin{equation}
{\rm Tr}U(1)_1= 36 ~~~~~~~{\rm and}~~~~~~~{\rm Tr}U(1)_3= -36.
\label{u1u3}
\end{equation}
The $E_6$ combination, given by,
\begin{equation}
U(1)_\zeta ~=~ U(1)_1+U(1)_2+U(1)_3~,
\label{uzeta}
\end{equation}
is anomaly free
and can be part of an unbroken $U(1)$ symmetry below the string scale.
The $SO(6)\times SO(4)$ observable gauge symmetry is
broken by the VEVs of the heavy Higgs fields ${\cal H}$ and
$\overline{\cal H}$ . The charges of these fields under the
Standard Model gauge group factors is given by:
\begin{align}
\overline{\cal H}({\bf\bar4},{\bf1},{\bf2})& \rightarrow u^c_H\left({\bf\bar3},
{\bf1},\frac 23\right)+d^c_H\left({\bf\bar 3},{\bf1},-\frac 13\right)+\nonumber\\
         &   ~~~~~~~~~~~~~~~~~~  {\overline {\cal N}}\left({\bf1},{\bf1},0\right)+
                             e^c_H\left({\bf1},{\bf1},-1\right)
                             \nonumber \\
{\cal H}\left({\bf4},{\bf1},{\bf2}\right) &
\rightarrow  u_H\left({\bf3},{\bf1},-\frac 23\right)+
d_H\left({\bf3},{\bf1},\frac 13\right)+\nonumber\\
         &  ~~~~~~~~~~~~~~~~~~   {\cal N}\left({\bf1},{\bf1},0\right)+
e_H\left({\bf1},{\bf1},1\right)\nonumber
\end{align}
The VEVs along the ${\cal N}$ and $\overline{\cal N}$ directions preserves
$N=1$ spacetime supersymmetry along a flat direction and 
leave the $U(1)$ combination
\begin{equation}
U(1)_{{Z}^\prime} ~=~
\frac{1}{5} (U(1)_C - U(1)_L) - U(1)_\zeta
~\notin~ SO(10),
\label{uzpwuzeta}
\end{equation}
unbroken below the string scale. This $U(1)$ combination is anomaly free provided that 
$U(1)_\zeta$ is anomaly free, as is the case in the string derived model \cite{Faraggi:2014ica}. 

We emphasise that this symmetry breaking pattern, and the $U(1)_{Z^\prime}$ combination
in eq. (\ref{uzpwuzeta}) is enforced
in the string derived model due the available scalar states in the string spectrum
to break the non--Abelian $SO(6)\times SO(4)$ gauge symmetry, and due
to a doublet--triplet missing partner mechanism \cite{Antoniadis:1988cm}
that gives heavy mass
to coloured scalar states that arise in the untwisted sector of the string
model \cite{Faraggi:2014ica}. Thus, the combination given in eq.
(\ref{uzpwuzeta}) is the extra $U(1)$ combination that can arise
in the string derived model with $E_6$ embedding of the charges.
Contrary to the situation in string inspired model, where any combination of 
$U(1)_C-U(1)_L$ and $U(1)_\zeta$ is possible, the $U(1)$ combination 
$U(1)_{Z^\prime}$ is the uniquely predicted combination in the string derived model. 

Anomaly cancellation of the $U(1)_{Z^\prime}$ gauge symmetry 
down to low scales, requires the vector--like leptons
$\{H_{\textrm{vl}}^i, {\bar H}_{\textrm{vl}}^i\}$, and quarks $\{D^i, \overline{D}^i\}$, 
that are obtained
from the vectorial $10$ multiplets of $SO(10)$, and the $SO(10)$
singlets $S^i$ in the $27$ representation of $E_6$.
The supermultiplet\footnote{Superfields are indicated with a hat symbol.}
states below the $SO(6)\times SO(4)$  breaking
scale are displayed schematically in Table~\ref{table27rot}.
The three right--handed
neutrino $N_L^i$ states obtain heavy mass at the $SU(2)_R$ breaking scale,
which produces the seesaw mechanism \cite{Faraggi:2018pit}.
The spectrum below the $SU(2)_R$ breaking scale is assumed to be
supersymmetric. We include in the spectrum
an additional pair of vector--like electroweak Higgs doublets,
that facilitate gauge coupling unification at the GUT scale.
This is justified in the string inspired models due to the string
doublet--triplet splitting mechanism \cite{Faraggi:1994cv, Faraggi:2001ry}.
We note from table~\ref{table27rot} that this extra Higgs pair is not chiral with respect
to $U(1)_{Z^\prime}$, contrary to the three pairs, $\{H_{\textrm{vl}}^i, {\bar H}_{\textrm{vl}}^i\}$,
that are chiral with respect to $U(1)_{Z^\prime}$. Therefore, mass terms for 
these three pairs can only be generated by the breaking of $U(1)_{Z^\prime}$, 
whereas the bilinear mass term for the extra vector--like Higgs pair can, 
in principle, be generated at a high scale.
The states $\phi$ and ${\bar\phi}$ are
exotic Wilsonian states \cite{Faraggi:2014ica, DelleRose:2017vvz}.
Additionally, the existence of light states $\zeta_i$,
that are neutral under the
$SU(3)_C\times SU(2)_L\times U(1)_Y\times U(1)_{Z^\prime}$ low
scale gauge group, is allowed.
The $U(1)_{Z^\prime}$
gauge symmetry can be broken at low scale by the VEV of the $SO(10)$ singlet fields 
$S_i$ and/or ${\phi_{1,2}}$.

\begin{table}[!ht]
\noindent
{\small
\begin{center}
{
\begin{tabular}{|l|cc|c|c|c|}
\hline
Field &$\hphantom{\times}SU(3)_C$&$\times SU(2)_L $
&${U(1)}_{Y}$&${U(1)}_{Z^\prime}$  \\
\hline
$\hat{Q}_L^i$&    $3$       &  $2$ &  $+\frac{1}{6}$   & $-\frac{2}{5}$   ~~  \\
$\hat{u}_L^i$&    ${\bar3}$ &  $1$ &  $-\frac{2}{3}$   & $-\frac{2}{5}$   ~~  \\
$\hat{d}_L^i$&    ${\bar3}$ &  $1$ &  $+\frac{1}{3}$   & $-\frac{4}{5}$  ~~  \\
$\hat{e}_L^i$&    $1$       &  $1$ &  $+1          $   & $-\frac{2}{5}$  ~~  \\
$\hat{L}_L^i$&    $1$       &  $2$ &  $-\frac{1}{2}$   & $-\frac{4}{5}$  ~~  \\
\hline
$\hat{D}^i$       & $3$     & $1$ & $-\frac{1}{3}$     & $+\frac{4}{5}$  ~~    \\
$\hat{{\bar D}}^i$& ${\bar3}$ & $1$ &  $+\frac{1}{3}$  &   $+\frac{6}{5}$  ~~    \\
$\hat{H}_{\textrm{vl}}^i$       & $1$       & $2$ &  $-\frac{1}{2}$   &  $+\frac{6}{5}$ ~~    \\
$\hat{{\bar H}}_{\textrm{vl}}^i$& $1$       & $2$ &  $+\frac{1}{2}$   &   $+\frac{4}{5}$   ~~  \\
\hline
$\hat{S}^i$       & $1$       & $1$ &  ~~$0$  &  $-2$       ~~   \\
\hline
$\hat{H}_1$         & $1$       & $2$ &  $-\frac{1}{2}$  &  $-\frac{4}{5}$  ~~    \\
$\hat{H}_2$  & $1$       & $2$ &  $+\frac{1}{2}$  &  $+\frac{4}{5}$  ~~    \\
\hline
$\hat{\phi}$       & $1$       & $1$ &  ~~$0$         & $-1$     ~~   \\
$\hat{\bar\phi}$       & $1$       & $1$ &  ~~$0$     & $+1$     ~~   \\
\hline
\hline
$\hat{\zeta}^i$       & $1$       & $1$ &  ~~$0$  &  ~~$0$       ~~   \\
\hline
\end{tabular}}
\end{center}
}
\caption{\label{table27rot}
Supermultiplet spectrum and
$SU(3)_C\times SU(2)_L\times U(1)_{Y}\times U(1)_{{Z}^\prime}$
quantum numbers, with $i=1,2,3$ for the three light
generations. The charges are displayed in the
normalisation used in free fermionic
heterotic--string models. }
\end{table}

\section{Phenomenological aspects of the model\label{pheno}}

In this section we illustrate details of the phenomenological analysis for the model described in the previous section whose charge assignment is given in Table~\ref{table27rot}. The analysis is based on previous work of two of the authors, published in refs.~\cite{Faraggi:2022emm,Faraggi:2015iaa,Coriano:2008wf,Coriano:2007ba} which is used as a reference for the notation. Some definitions are also imported for consistency.

\subsection{Neutral gauge bosons sector}
We start by analyzing the neutral gauge bosons sector. 
The covariant derivative relative to the low-scale gauge group $SU(3)_C \times SU(2) \times U(1)_Y \times U(1)_{Z'}$ is defined as 
\begin{equation}
{\cal D}_{\mu} = \partial_{\mu} + i g_s A_{\mu}^{a} T^a + i g_2 W_{\mu}^a \tau^a + i g_Y \frac{Y}{2} A^Y_{\mu} + i g_{Z'} \frac{z}{2} B_{\mu} \label{eq::covderivative1}
\end{equation}
where $g_2$ is the $SU(2)$ coupling and $W_{\mu}^a$ and $\tau^a$ are the $SU(2)$ gauge fields and generators respectively; $g_Y$ is the $U(1)_Y$ coupling, while $A^Y_{\mu}$ is the gauge field and $Y$ is the hypercharge.
$g_{Z'}$ is the $U(1)_{Z'}$ coupling and $B_{\mu}$ and $z$ are the gauge field and charge of the $U(1)_{Z'}$ respectively. $g_s$ is the strong coupling constant, and $A^a$ and $T^a$ are the $SU(3)_{C}$ gauge fields and generators respectively.
 
The lagrangian of the Higgs sector is given 
\begin{equation}
    \mathcal{L} = \frac{1}{2} \left| {\cal D}_\mu H_1 \right| ^2 + \frac{1}{2} \left| {\cal D}_\mu H_2 \right| ^2 + \frac{1}{2} \sum_{i}\left| {\cal D}_\mu S_i \right| ^2
\end{equation}
where the scalar components of the supermultiplets which give the two Higgs doublets are represented by the $H_1$ and $H_2$ fields, that have hypercharge $Y_{H1}=-1$, $Y_{H2}=1$ and $U(1)_{Z'}$ charges $z_{H1}$ and $z_{H2}$. The singlets are the $S_i$ fields, and as in~\cite{Faraggi:2022emm} we consider only one $S$ field for simplicity. 
The hypercharge of $S$ is $Y_{S}=0$ and its charge under $U(1)_{Z'}$ is $z_{s}$. 

\begin{figure}
    \centering
    \includegraphics{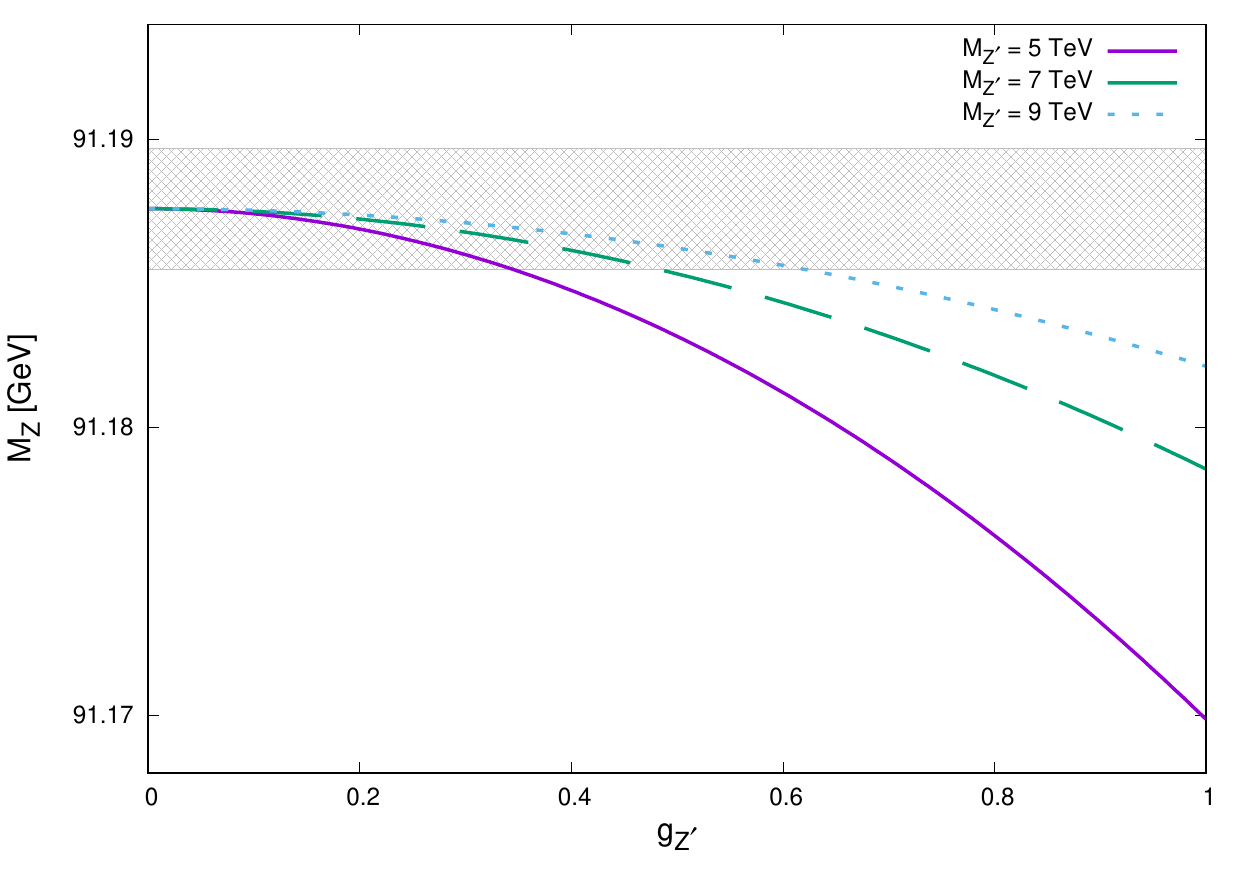}
    \caption{$Z'$-induced corrections on the SM $Z$ boson mass as a function of $g_{Z'}$ for fixed values of $M_{Z'}$. The hatched band represents the current measured $M_Z$ uncertainty as reported in the PDG~\cite{ParticleDataGroup:2022pth}.}
    \label{fig:zmass}
\end{figure}

The Higgs fields are parametrized as 
\begin{equation}
H_1 = \begin{pmatrix} \text{Re} H_1^0 + i \text{Im} H_1^0 \\ \text{Re} H_1^- + i \text{Im} H_1^- \end{pmatrix}, 
~~~~~
H_2 = \begin{pmatrix} \text{Re} H_2^+ + i \text{Im} H_2^+ \\ \text{Re} H_2^0 + i \text{Im} H_2^0 \end{pmatrix}, 
~~~~~
S = \text{Re}S + i \text{Im} S,
\end{equation}
and the vacuum expectation values (VEVs) are defined as
\begin{equation}
    \langle H_1 \rangle = \begin{pmatrix} v_1 \\ 0 \end{pmatrix}, ~~
    \langle H_2 \rangle = \begin{pmatrix} 0 \\ v_2 \end{pmatrix}, ~~
    \langle S \rangle = v_s.
\end{equation}
Expanding the Higgs lagrangian and collecting quadratic terms, we obtain the mass matrix in the $(W^3_\mu,A^Y_\mu,B_\mu)$ basis as
\begin{equation}
    M^2 = \frac{1}{4} \begin{pmatrix} g_2^2v^2 & -g_2g_Yv^2 & g_2x_z \\ -g_2g_Yv^2 & g_Y^2v^2 & -g_Yx_z \\ g_2x_z & -g_Yx_z & N_z\end{pmatrix}
    \label{eq::massmatrix}
\end{equation}
where for convenience we have defined the following quantities 
\begin{equation}
    v^2=v_1^2+v_2^2, ~~~
    x_z=g_{Z'}(z_{H1}v_1^2-z_{H2}v_2^2), ~~~
    N_z=g_{Z'}^2(z_{H1}^2v_1^2+z_{H2}^2v_2^2+z_s^2v_s^2).
\end{equation}
The zero eigenvalue of the mass matrix corresponds to the photon. The other two eigenvalues correspond to the square mass of the $Z$ and $Z'$ physical gauge bosons 
\begin{equation}
\begin{aligned}
    M_Z^2 = \frac{1}{8}\left( g^2v^2+N_z-\sqrt{(N_z-g^2v^2)^2+4g^2x_z^2}\right), \\
    M_{Z'}^2 = \frac{1}{8}\left( g^2v^2+N_z+\sqrt{(N_z-g^2v^2)^2+4g^2x_z^2}\right)
\end{aligned}
\label{ZZp-masses}
\end{equation}
where $g^2=g_2^2+g_Y^2$. From Eq.~\ref{ZZp-masses} we see that the presence of the $Z'$ induces corrections on the mass of the SM $Z$ boson. In Fig.~\ref{fig:zmass} we show the impact of these corrections on the SM $Z$ boson mass within the current uncertainties reported in the PDG~\cite{ParticleDataGroup:2022pth} as a function of $g_{Z'}$ for different values of $M_{Z'}$.  
We note that for $g_{Z'}\lesssim 0.4$ deviations are well within the current uncertainties. 

In the limit where $v_s$ is very large, the $Z'$ mass decouples and one obtains simplified expressions for the gauge boson masses,
\begin{equation}
    M_Z^2 = \frac{1}{4}g^2v^2\left[ 1+\mathcal O(1/v_s^2)\right]\,,~~~ 
    M_{Z'}^2 = \frac{1}{4}N_z\left[ 1+\mathcal O(1/v_s^2)\right]
\end{equation}

From the mass matrix diagonalization we obtain the normalized orthogonal eigenvector matrix which is used to rotate the gauge field basis to the physical field basis. The exact form of the elements of this matrix is given in Appendix~\ref{appendix:rot}. When $v_s$ is large, the rotation matrix can be simplified as
\begin{equation}
    \label{eq::rotmatrix}
    \begin{pmatrix} W^3_\mu \\ A^Y_{\mu} \\ B_{\mu} \end{pmatrix} = 
    \begin{pmatrix} \sin\theta_W && \cos\theta_W \cos\delta && \cos\theta_W \sin\delta \\ \cos\theta_W && -\sin\theta_W \cos\delta && -\sin\theta_W \sin\delta \\ 0 && -\sin\delta && \cos\delta
    \end{pmatrix}
    \begin{pmatrix} A^\gamma_\mu \\ Z_\mu \\ Z'_\mu \end{pmatrix}
\end{equation}
where $\theta_W$ is the Weinberg angle, $\sin\theta_W = g_2 / g$ and $\cos\theta_W = g_Y / g$ as in the Standard Model, and the small mixing between the $Z$ and $Z'$ is parametrized by the angle $\delta$, defined by 
\begin{equation}
    \sin\delta = \frac{gx_z}{N_z-g^2v^2} + \mathcal O\left(\frac{1}{v_s^2}\right)
\end{equation}
To simplify the notation, we define a perturbative parameter $\varepsilon = \sin\delta$. The rotation matrix in Eq.~\ref{eq::rotmatrix} to first order in $\varepsilon$ can now be written as 
\begin{equation}
    O_{gauge} = 
    \begin{pmatrix} \sin\theta_W & \cos\theta_W & \varepsilon \cos\theta_W \\ \cos\theta_W & -\sin\theta_W & -\varepsilon \sin\theta_W \\ 0 & -\varepsilon & 1 \end{pmatrix}
\end{equation}
According to the parameter space explored in this work, $\varepsilon$ is typically $10^{-4}\lesssim \varepsilon \lesssim 10^{-3}$. After fields are rotated to the physical basis, the covariant derivative for the neutral current (NC) sector of Eq.~\ref{eq::covderivative1} in terms of the mass eigenstate fields is given by
\begin{equation}
\begin{split}
{\cal D}_{\mu}^{NC} & = \partial_{\mu} + i \left( g_2 \sin\theta_W \tau^3 + g_Y \cos\theta_W\frac{Y}{2} \right) A^{\gamma}_{\mu} \\
& \quad + i \left( g_2 \cos\theta_W \tau^3 - g_Y \sin\theta_W \frac{Y}{2} - \varepsilon g_{Z'} \frac{z}{2} \right) Z_{\mu} \\ 
& \quad + i \left( \varepsilon g_2 \cos\theta_W \tau^3 - \varepsilon g_Y \sin\theta_W \frac{Y}{2} + g_{Z'} \frac{z}{2} \right) Z^{\prime}_{\mu}
\end{split}
\end{equation}
The neutral current contributions in the Lagrangian, that are of the form $\bar{\psi} \gamma^\mu c^Z \psi Z_\mu$ and $\bar{\psi} \gamma^\mu c^{Z'} \psi Z'_\mu$, can be rewritten using the relations $g_Y\cos\theta_W=g_2 \sin\theta_W$, $Q=\tau^3+\frac{Y}{2}$, $e=g_2\sin\theta_W$
\begin{equation}
\begin{aligned}
& c^{Z} = g (\tau^{3}-\sin^2\theta_WQ) - \varepsilon g_{Z'} \frac{z}{2} \\
& c^{Z'} = \varepsilon g(\tau^{3}-\sin^2\theta_WQ)+g_{Z'}\frac{z}{2}
\end{aligned}
\end{equation}
where the charges $\tau^3, Q, z$ correspond to the fermion and $\varepsilon$ is the (small) mixing parameter.     
Using this notation, the interactions between fermions and electroweak neutral gauge bosons are expressed in terms of the vector and axial vector couplings, i.e., $\bar{\psi} \gamma^\mu \frac{g}{4} \left( g_V^Z - g_A^Z \gamma^5 \right) \psi Z_\mu$ and $\bar{\psi} \gamma^\mu \frac{g}{4} \left( g_V^{Z'} - g_A^{Z'} \gamma^5 \right) \psi Z'_\mu$. These couplings are explicitly given as
\begin{equation}
\begin{split}
    \frac{g}{4} g_V^Z &= \frac{1}{2} \left[ g \left( \tau^3_L - 2 \sin^2 \theta_W Q \right) - \varepsilon g_{Z'} \left( \frac{z_L}{2} + \frac{z_R}{2} \right) \right] \\
    \frac{g}{4} g_A^Z &= \frac{1}{2} \left[ g \tau^3_L - \varepsilon g_{Z'} \left( \frac{z_L}{2} - \frac{z_R}{2} \right) \right] \\
    \frac{g}{4} g_V^{Z'} &= \frac{1}{2} \left[ \varepsilon g \left( \tau^3_L - 2 \sin^2 \theta_W Q \right) + g_{Z'}\left( \frac{z_L}{2} + \frac{z_R}{2} \right) \right] \\
    \frac{g}{4} g_A^{Z'} &= \frac{1}{2} \left[ \varepsilon g \tau^3_L + g_{Z'} \left( \frac{z_L}{2} - \frac{z_R}{2} \right) \right] \\
\end{split}
\end{equation}
where $\tau^3_L, z_{L/R}$ refer to left/right-handed fermions.

\subsection{Kinetic mixing}

In this section we generalize the results obtained in the previous section to the case of kinetic-mixing interactions. The impact of kinetic mixing is assessed in Sec.~\ref{numerical}, where applications to hadron collider phenomenology are discussed.

Since the low-energy gauge group includes two abelian groups, the Lagrangian for the kinetic terms can include a contribution which directly couples the $A^{Y}$ and $B$ fields and does not break either $U(1)_Y$ or $U(1)_{Z'}$ gauge invariance. This term corresponds to gauge kinetic mixing and is of the form
\begin{equation}
\Delta\mathcal L_{kin} = -\chi A^Y_{\mu\nu} B^{\mu\nu}
\end{equation}
where $A^Y_{\mu\nu}$ and $B_{\mu\nu}$ are the field-strength tensors for the $U(1)_Y$ and $U(1)_{Z'}$ symmetry groups respectively, and $\chi$ is a parameter such that $|\chi|< 1$. This renormalizable operator plays an interesting role in the context of our model because it can be produced at scales much higher than the $Z'$ breaking scale and can affect the couplings of $Z'$s at the TeV scale. It is therefore interesting to study the interplay between the gauge kinetic-mixing parameter and coupling of the $Z'$.
The expression for the electroweak NC covariant derivative is generalized as %
\begin{equation}
{\cal D}_{\mu}^{NC} = \partial_{\mu} + ig_2W^3_{\mu}\tau^3 + iQ^TGA
\end{equation}
where $Q^T = (Y/2, z/2)$ is a vector containing the charges for the two $U(1)$ groups, $G$ is the mixing matrix which is defined as 
\begin{equation}
G = \begin{pmatrix} g_{AA} && g_{AB} \\ g_{BA} && g_{BB}
\end{pmatrix}
\end{equation}
and contains the gauge couplings, and $A=(A^Y_\mu,B_\mu)^T$ contains the $U(1)$ gauge fields. The off-diagonal elements $g_{AB}$ and $g_{BA}$ contain the kinetic mixing between the $U(1)$ fields.

It is convenient to perform the rotation $A\rightarrow A' = RA$, $G \rightarrow G' = GR^{-1}$ where $R$ is an orthogonal $2\times2$ matrix, and choose $g_{BA}'=0$ in order to eliminate the $z_sg_{BA}A^Y_\mu$ term in ${\cal D}_\mu^{NC}$. This way, the Higgs singlet does not affect the $SU(2)\times U(1)_Y$ sector in the mass matrix. In the new basis, the mixing matrix reads
\begin{equation}
    G'=
    \begin{pmatrix} g_Y && g_{KM} \\ 0 && g_{Z'}
    \end{pmatrix}
\end{equation}
where
\begin{equation}
\begin{aligned}
    &g_Y = (g_{AA}g_{BB}-g_{AB}g_{BA}) / \sqrt{g_{BB}^2+g_{AB}^2} \\
    &g_{Z'} = \sqrt{g_{BB}^2+g_{BA}^2} \\
    &g_{KM} = (g_{AA}g_{BA}+g_{AB}g_{BB}) / \sqrt{g_{BB}^2+g_{BA}^2} 
\end{aligned}
\end{equation}
The electroweak NC covariant derivative becomes
\begin{equation}
    {\cal D}^{NC}_\mu = \partial_\mu + ig_2W^3_\mu \tau^3 + ig_Y\frac{Y}{2}A^{Y\prime}_{\mu} + i\left( g_{Z'}\frac{z}{2} + g_{KM}\frac{Y}{2}\right) B_{\mu}'
\end{equation}
where $A^{Y\prime}_{\mu}$ and $B_\mu'$ are the rotated $U(1)$ gauge fields. 
Therefore, the inclusion of gauge kinetic mixing can be studied by using a single parameter $g_{KM}$. The mass matrix in the $(W^3_\mu,A^{Y\prime}_\mu,B_\mu')$ basis is now 
\begin{equation}
    \frac{1}{4}
    \begin{pmatrix}
    g_2^2v^2 && -g_2g_Yv^2 && g_2x_z' \\ -g_2g_Yv^2 && g_Y^2v^2 && -g_Yx_z' \\ g_2x_z' && -g_Yx_z' && N_z'
    \end{pmatrix}
\end{equation}
where $x_z$ and $N_z$ have been replaced by their primed expressions given by
\begin{equation} \label{eq:xz_km}
\begin{aligned}
    &x_z' = g_{Z'} (z_{H1}v_1^2 - z_{H2}v_2^2) - g_{KM}v^2 \\
    &N_z' = (g_{Z'}z_{H1} - g_{KM})^2v_1^2 + (g_{Z'}z_{H2} + g_{KM})^2v_2^2 + g_{Z'}^2z_s^2v_s^2
\end{aligned}
\end{equation}
The calculation for the gauge boson masses and the rotation matrix to the physical gauge boson basis proceeds identically to the previous section, with the replacements $x_z \rightarrow x_z'$ and $N_z \rightarrow N_z'$. With these modifications, the general form of the fermionic couplings to the $Z$ and $Z'$ in presence of kinetic mixing is
\begin{equation}
\begin{aligned}
    &c^Z = g (\tau^3 - \sin^2 \theta_W Q) - \varepsilon' \left(g_{Z'} \frac{z}{2} + g_{KM} \frac{Y}{2}\right) \\
    &c^{Z'} = \varepsilon' g (\tau^3 - \sin^2 \theta_W Q) + g_{Z'}\frac{z}{2} + g_{KM} \frac{Y}{2}\,,
\end{aligned}
\end{equation}
where the perturbative parameter from the previous section is modified as
\begin{equation}
\varepsilon' = \frac{gx_z'}{N_z'-g^2v^2} + \mathcal O\left(\frac{1}{v_s^2}\right)
\label{epsprime}
\end{equation}
The vector and axial vector couplings of the gauge bosons to the fermions are now
\begin{equation}
\begin{split}
    \frac{g}{4} g_V^Z &= \frac{1}{2} \left[ g \left( \tau^3_L - 2 \sin^2 \theta_W Q \right) - \varepsilon' g_{Z'} \left( \frac{z_L}{2} + \frac{z_R}{2} \right) - \varepsilon' g_{KM} \left( \frac{Y_L}{2} + \frac{Y_R}{2} \right) \right] \\
    \frac{g}{4} g_A^Z &= \frac{1}{2} \left[ g \tau^3_L - \varepsilon' g_{Z'} \left( \frac{z_L}{2} - \frac{z_R}{2} \right) - \varepsilon' g_{KM} \left( \frac{Y_L}{2} - \frac{Y_R}{2} \right) \right] \\
    \frac{g}{4} g_V^{Z'} &= \frac{1}{2} \left[ \varepsilon' g \left( \tau^3_L - 2 \sin^2 \theta_W Q \right) + g_{Z'}\left( \frac{z_L}{2} + \frac{z_R}{2} \right) + g_{KM} \left( \frac{Y_L}{2} + \frac{Y_R}{2} \right) \right] \\
    \frac{g}{4} g_A^{Z'} &= \frac{1}{2} \left[ \varepsilon' g \tau^3_L + g_{Z'} \left( \frac{z_L}{2} - \frac{z_R}{2} \right) + g_{KM} \left( \frac{Y_L}{2} - \frac{Y_R}{2} \right) \right] \\
\end{split}
\label{gkm-Zp-couplings}
\end{equation}

\section{Hadron Collider Phenomenology Applications\label{numerical}}

In this section we explore the parameter space of the model and perform a detailed analysis for proton-proton collisions at the LHC using Drell-Yan (DY) kinematic distributions calculated at the next-to-next-to-leading order (NNLO) in the strong coupling constant $\alpha_s$ of Quantum Chromodynamics (QCD). 
The contribution of electroweak corrections is not considered here and their impact will be analyzed in a forthcoming work. We explore the sensitivity of the forward-backward asymmetry $(A_{FB})$ distribution to the $Z'$ and its parameters. 
In particular, we use $A_{FB}$ distributions to study the interplay between the gauge kinetic-mixing parameter and coupling of the $Z'$. It has been pointed out in several works (e.g., see~\cite{London:1986dk,Rosner:1986cv,Hewett:1988xc,Cvetic:1995zs,Rosner:1995ft,Dittmar:1996my,Bodek:2001xk,CDF:1997wdd,Accomando:2015cfa,Fiaschi:2022wgl,Ball:2022qtp} and references therein) that $A_{FB}$ distributions are very sensitive to SM deviations in the electroweak sector, and in particular to the presence of extra gauge vector bosons. Moreover, the ATLAS, LHCb, and CMS collaborations at the LHC have performed high-precision measurements of $A_{FB}$ distributions at 7, 8, and 13 TeV collision energy respectively~\cite{ATLAS:2015ihy,LHCb:2015jyu,CMS:2022uul}. The CMS collaboration~\cite{CMS:2022uul} has recently set lower limits on the mass of additional gauge bosons from sequential SM extensions at around 4.4 TeV.

Our theory predictions for the DY cross section are complemented by the calculation of uncertainties induced by parton distribution functions (PDFs) in the proton, as well as scale uncertainties for some of the distributions. PDFs represent one of the major sources of uncertainty in cross section calculations and complicate model validation and discrimination.    
The expression for the DY cross section in QCD factorization can be written as 
\begin{eqnarray} 
d\sigma = \sum_{ij}\int_0^1 d x_1 d x_2 
f_{i}^{H_1}(x_1,\mu_F^2) 
f_{j}^{H_2}(x_2,\mu_F^2) 
d\hat{\sigma}_{ij\rightarrow l \bar{l}}
(x_1,x_2;\alpha_s(\mu_R^2),\mu_R^2,\mu_F^2) 
+ {\cal O}\left(\frac{\Lambda_{\textrm{QCD}}^2}{Q^2} \right)
\end{eqnarray}
where $f_i^{H_1}$ and $f_j^{H_2}$ are the proton PDFs, which depend on the longitudinal momentum fraction $x_1$ and $x_2$ of parton $i$ and $j$ respectively,
and on the factorization scale $\mu_F$. $d\hat{\sigma}$ is the hard scattering cross section, which is perturbatively calculable in QCD, $\mu_R$ is the renormalization scale, and $\left( \Lambda_{\textrm{QCD}}^2/Q^2 \right)$ represents power-suppressed contributions where $\Lambda_{\textrm{QCD}}$ is the QCD scale.

The calculation of the differential distributions at NNLO in QCD which we present in this work has been performed by using an amended version of the \texttt{MCFM-v9.0} computer program~\cite{Campbell:1999ah,Campbell:2011bn,Campbell:2015qma,Boughezal:2016wmq,Campbell:2019dru}, which has been modified to incorporate the string-derived model with charge assignments in Table~\ref{table27rot}, the $Z'$ contribution, as well as the interference terms. We validated this implementation against other computer programs such as \texttt{DY-Turbo}~\cite{Camarda:2019zyx} and \texttt{FEWZ}~\cite{Gavin:2010az,Li:2012wna} and found agreement within 1\%. 
Our results are presented at $\sqrt{S} = $ 13 TeV of center-of-mass energy and we used the CT18NNLO~\cite{Hou:2019efy} PDFs with conservative uncertainties evaluated at the 90\% confidence level (CL). As a case study, we have chosen a string-derived $Z'$ with $M_{Z'}=$ 5 TeV. The impact of other recent PDF determinations~\cite{Alekhin:2017kpj,Bailey:2020ooq,NNPDF:2021njg} on $A_{FB}$ distributions in the context of extra resonance searches is studied in refs.~\cite{Fiaschi:2022wgl,Ball:2022qtp}. Scale uncertainties are obtained by using the 7-point variation, that is, varying $\mu_R$ and $\mu_F$ up and down independently by a factor of 2, and then taking the envelope.  

For the scope of this analysis, it is sufficient to include in the total decay rate of the $Z'$ entering the cross section calculation, only the major decay channels which we report in the expression below
\begin{eqnarray}
    \Gamma_{Z'} = \sum_f \Gamma_{Z'\rightarrow f\bar{f}} + \Gamma_{Z'\rightarrow W^+W^-} + \Gamma_{Z'\rightarrow H^+H^-}\,,
\end{eqnarray}
where $f$ runs over the quarks and leptons, $W^\pm$ are the charged gauge vector bosons, and $H^\pm$ are the Higgs bosons of the charged sector. The expression for the $\Gamma_{Z'\rightarrow H^+H^-}$ channel in presence of gauge kinetic mixing is given in Appendix~\ref{appendix:Zp-rates}. The calculation of the partial rates relative to the other channels proceeds similarly to that in refs.~\cite{Faraggi:2022emm,Coriano:2008wf}: in the expression for the $Z'$ decay rate in the fermion channel, the vector and axial vector couplings are replaced by those in Eq.~\ref{gkm-Zp-couplings}, while in the expression for the rate in the $W^+$ $W^-$ channel $\varepsilon$ is replaced by $\varepsilon'$ from Eq.~\ref{epsprime}.

\subsection{Kinematic distribution results \label{DY-inv-mass}}

\begin{figure}
    \includegraphics[width=9cm]{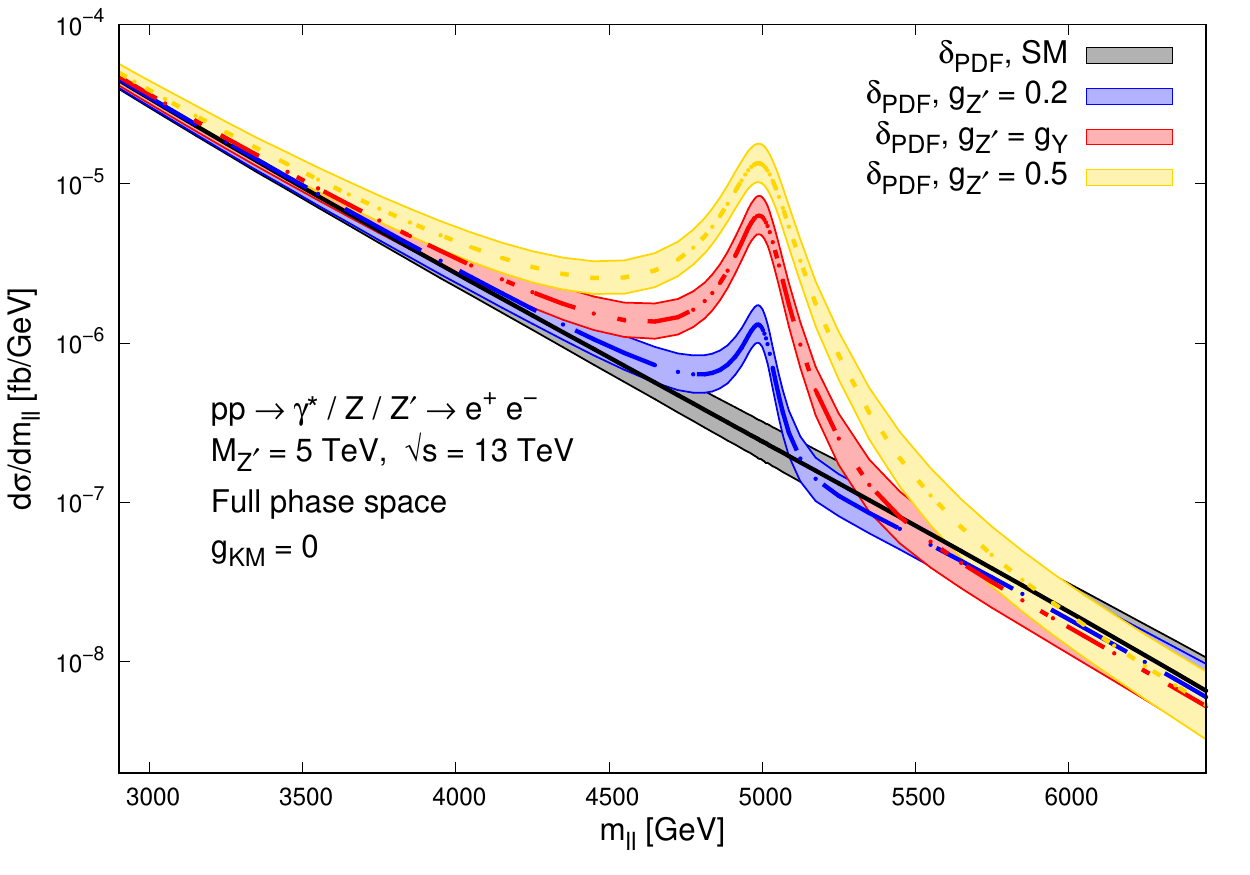}
    \includegraphics[width=9cm]{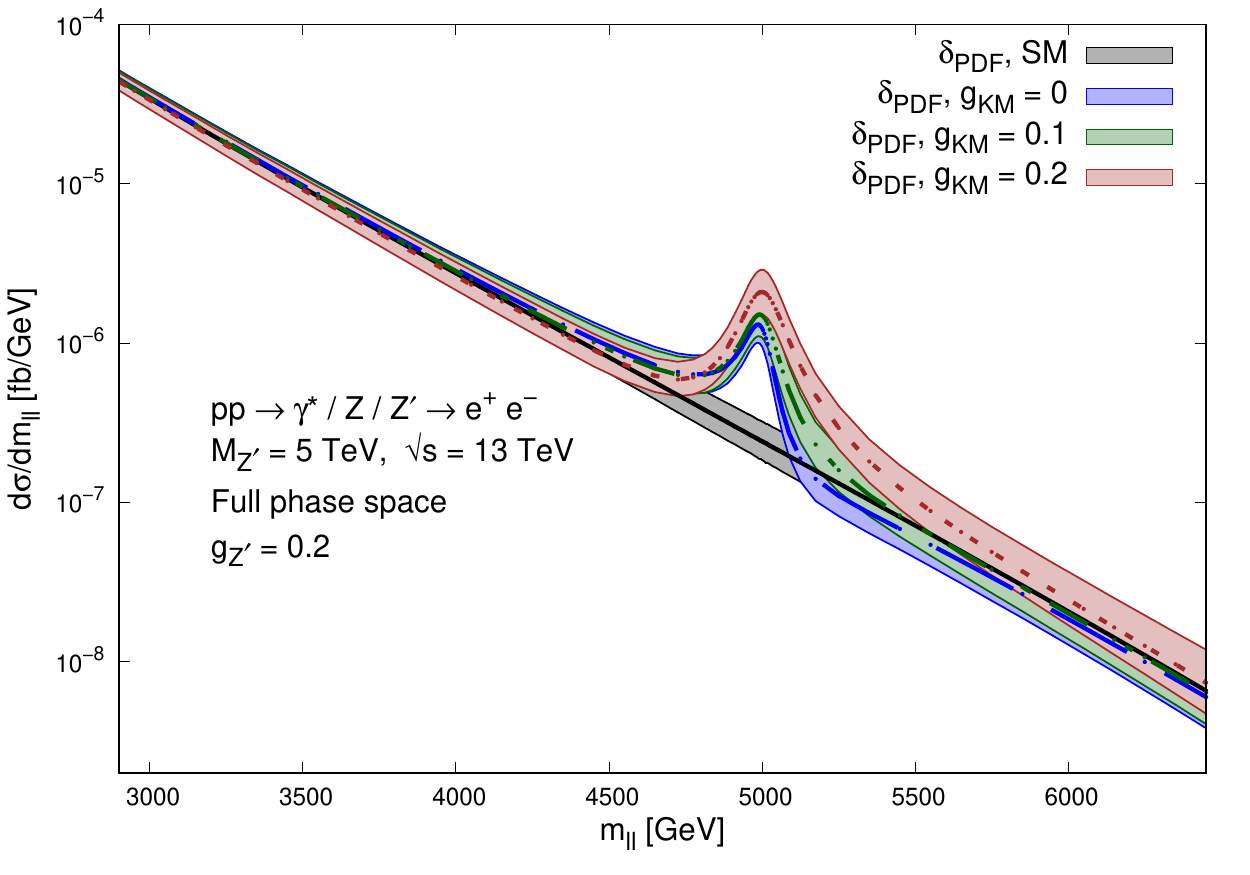}
    \includegraphics[width=9cm]{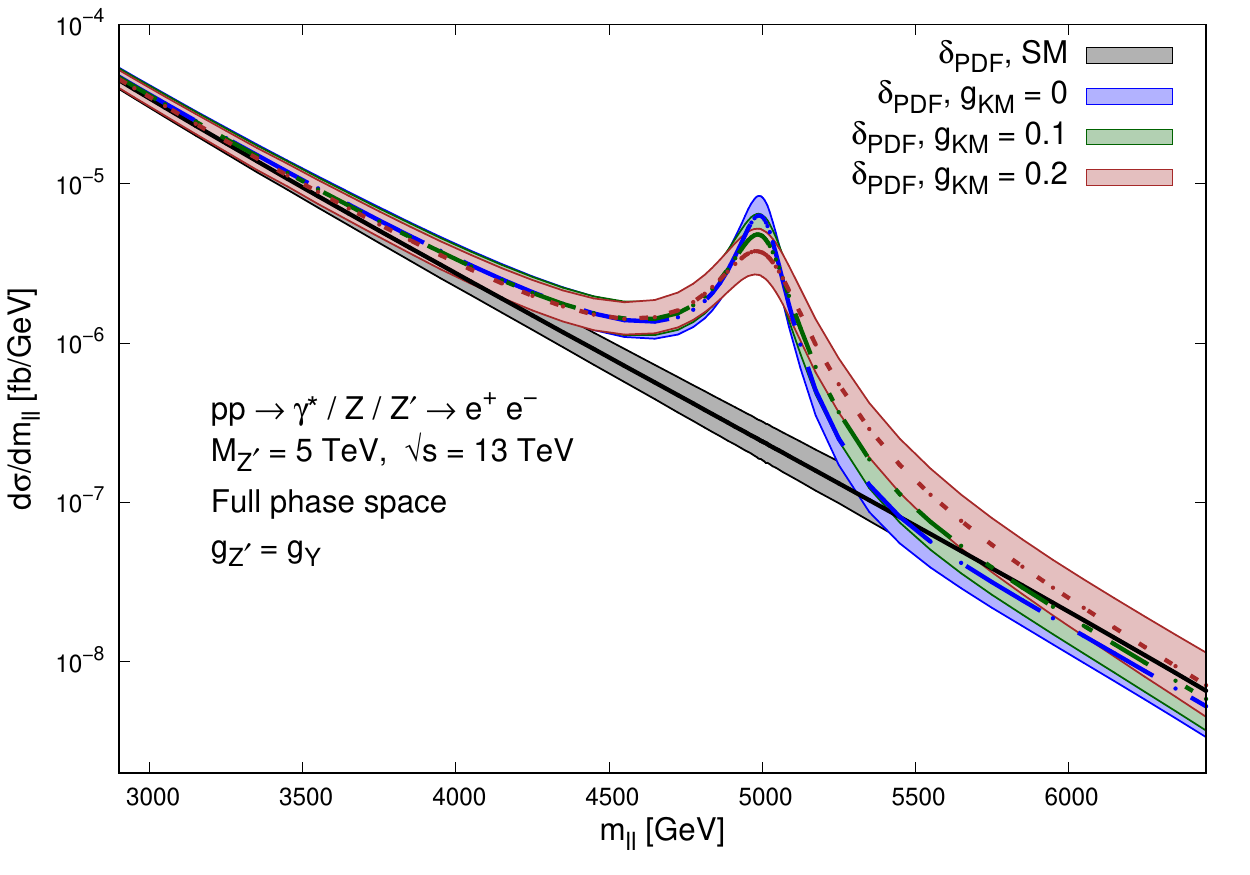}
    \caption{$Z'$ invariant mass distributions for varying $g_{Z'}$ and $g_{KM}$. The error bands represent the CT18NNLO induced PDF uncertainty on the cross section at 90\% CL. Central predictions are represented by lines with different dashing.}
    \label{inv-mass}
\end{figure}

In Fig.~\ref{inv-mass} we show the string-derived $Z'$ results for the dilepton invariant mass ($m_{ll}$) DY spectrum at the LHC 13 TeV in the electron channel, for a $Z'$ of mass $M_{Z'}$ = 5 TeV and different values of the coupling $g_{Z'}$ and kinetic-mixing parameter $g_{KM}$. These theory predictions are compared to the SM  
and the cross sections are calculated in the full phase space. The error bands represent the CT18NNLO induced PDF uncertainty evaluated at the 90\% CL. Central predictions are represented by lines with different dashing. We observe how the interplay between $g_{KM}$ and $g_{Z'}$ modifies the shape of the resonance and in particular the width for $g_{Z'} = 0.2$ and $g_{Z'} = g_Y$, when the strength of $g_{KM}$ is varied. The inclusion of kinetic mixing has a significant effect on the width of the resonance in the invariant mass distribution. 

\begin{figure}
\includegraphics[width=8.6cm]{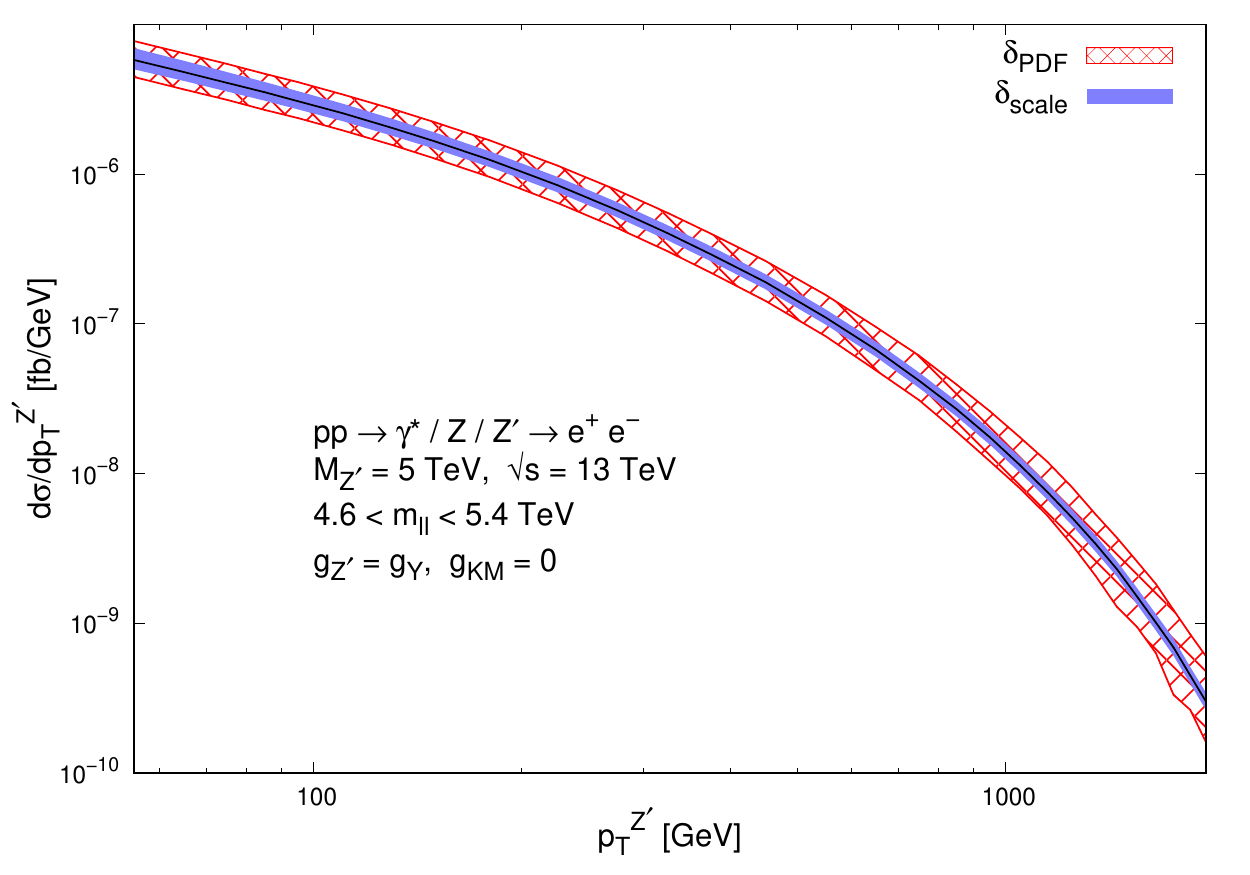}
\includegraphics[width=8.6cm]{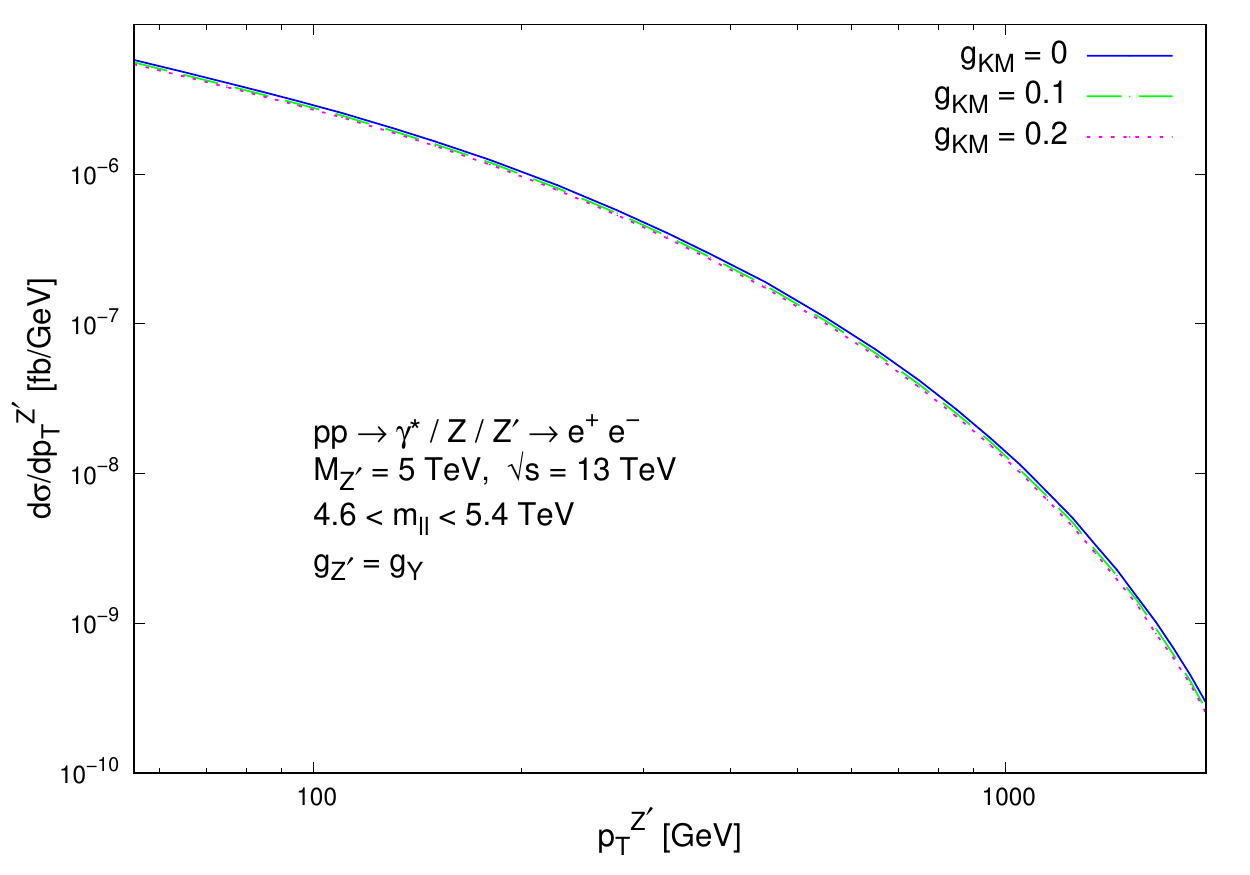}
\includegraphics[width=8.6cm]{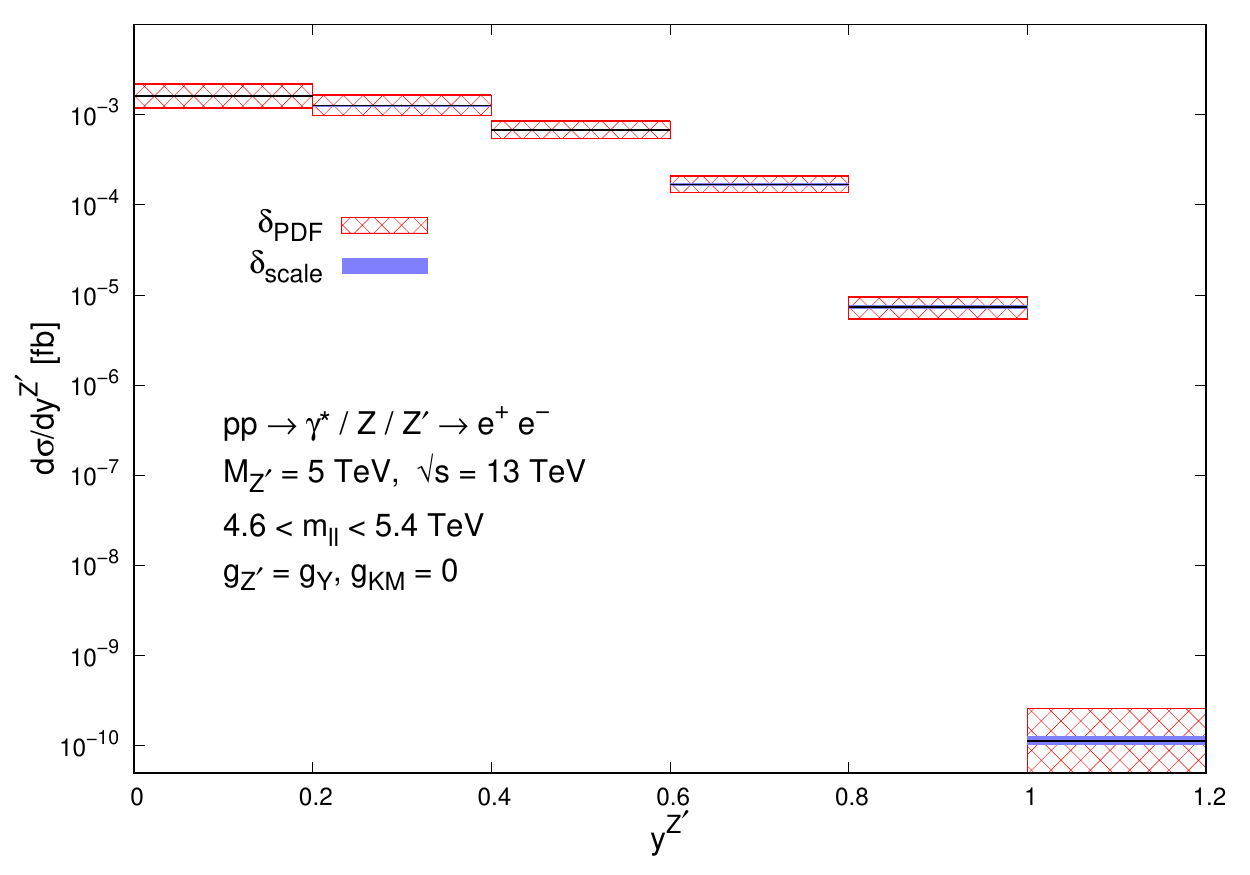}
\includegraphics[width=8.6cm]{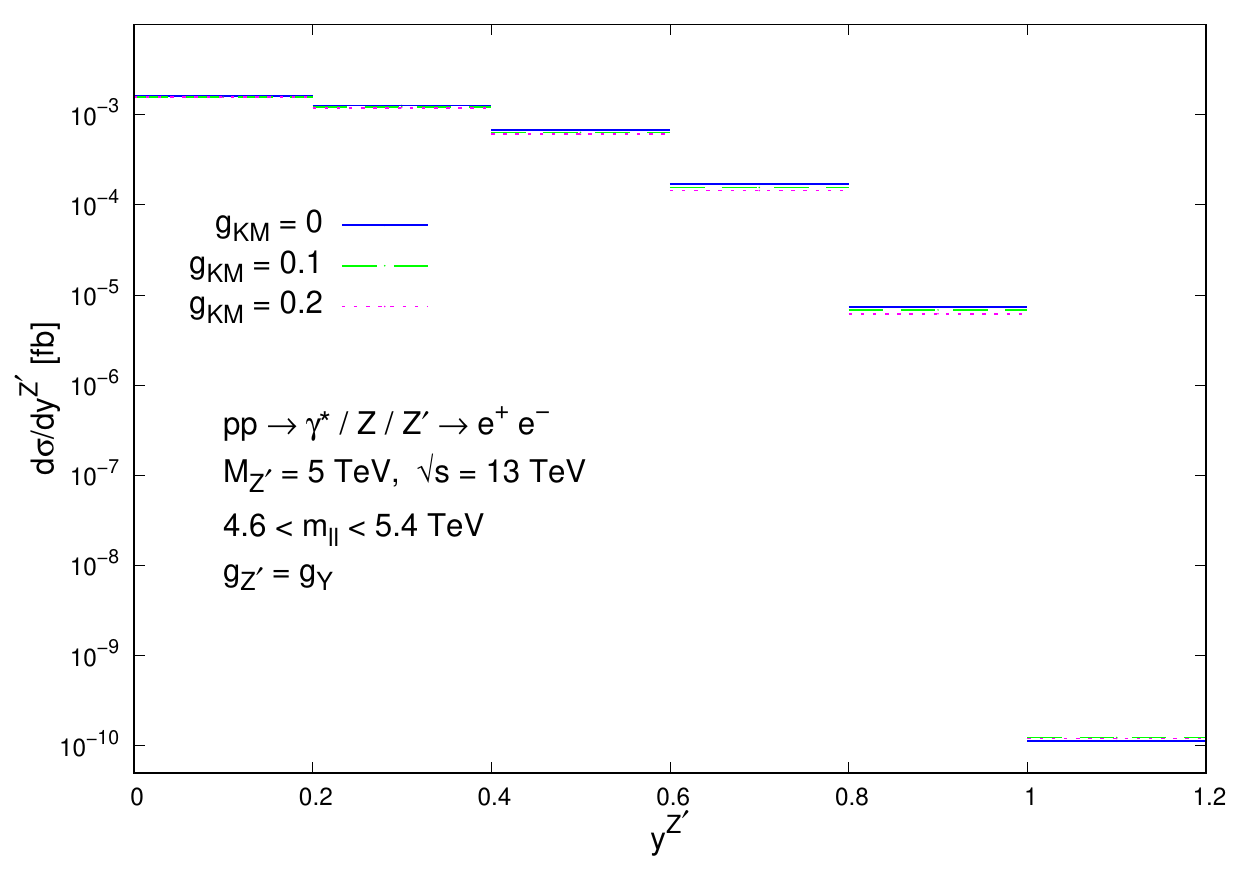}
\caption{Transverse momentum $p_T$ (upper) and rapidity $y_{Z'}$ of the $Z'$ (lower). The CT18NNLO PDF uncertainty is shown as a hatched red band and scale uncertainty is shown as a solid blue band. Central predictions are represented by a solid black line.}
\label{fig:pt34-y34}
\end{figure}

It is interesting to explore the impact of varying these parameters on more differential observables such as the $Z'$ transverse momentum $p_T$ spectrum and the $Z'$ rapidity distribution $y_{Z'}$, which we illustrate in Fig.~\ref{fig:pt34-y34}. Here, the invariant mass of the final-state dilepton pair is chosen to be $4.6 < m_{l\bar{l}}<5.4$ TeV. In the left-column insets, where $g_{KM}=0$, PDF uncertainties are represented by red-hatched bands while scale uncertainties are represented by blue bands. In the right-column insets we illustrate for clarity the same distributions with no uncertainties but with $g_{KM}>0$. We observe that kinetic mixing has negligible impact on these two distributions for this choice of the parameters in the kinematic region $4.6 < m_{l\bar{l}}<5.4$ TeV. Again, we observe that PDF uncertainty dominates.     

\begin{figure}
\includegraphics[width=9cm]{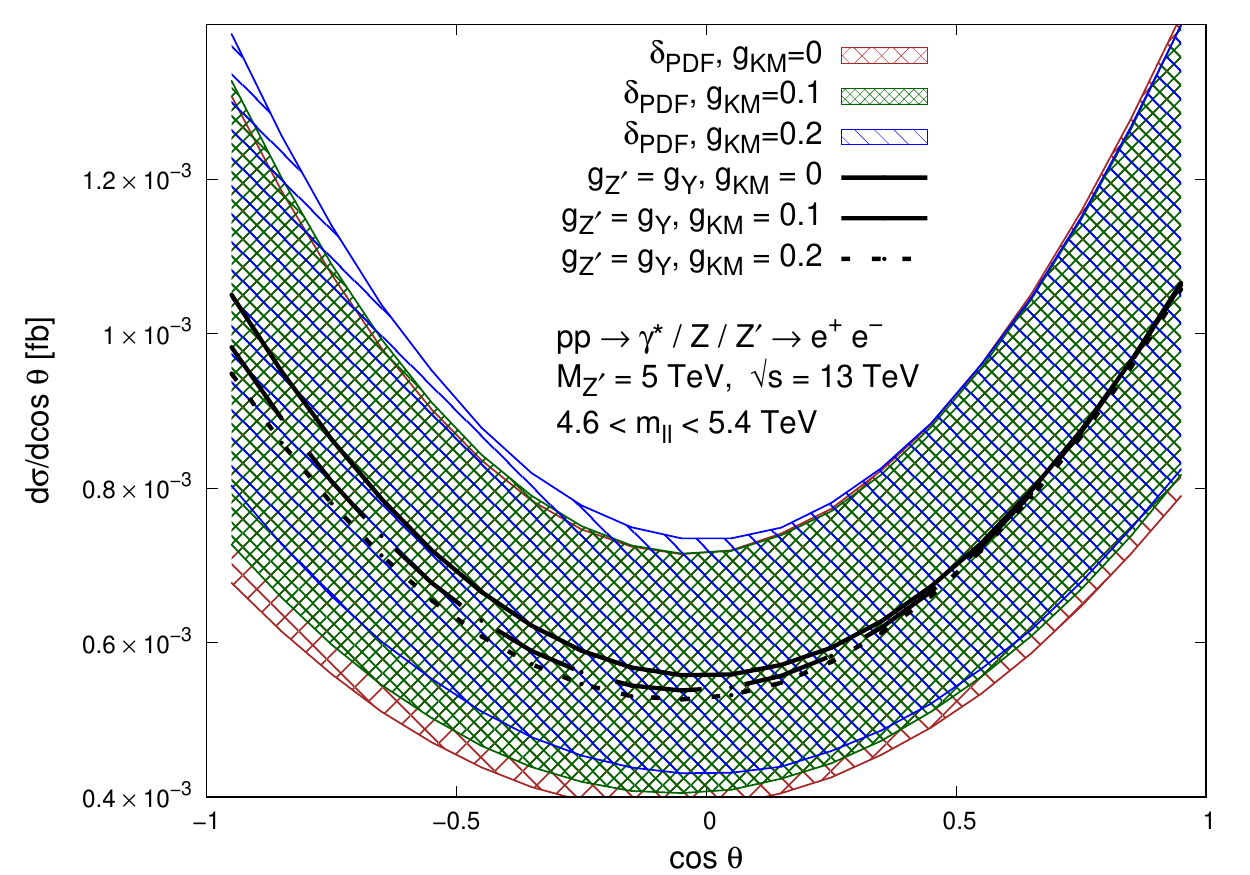}
\includegraphics[width=9cm]{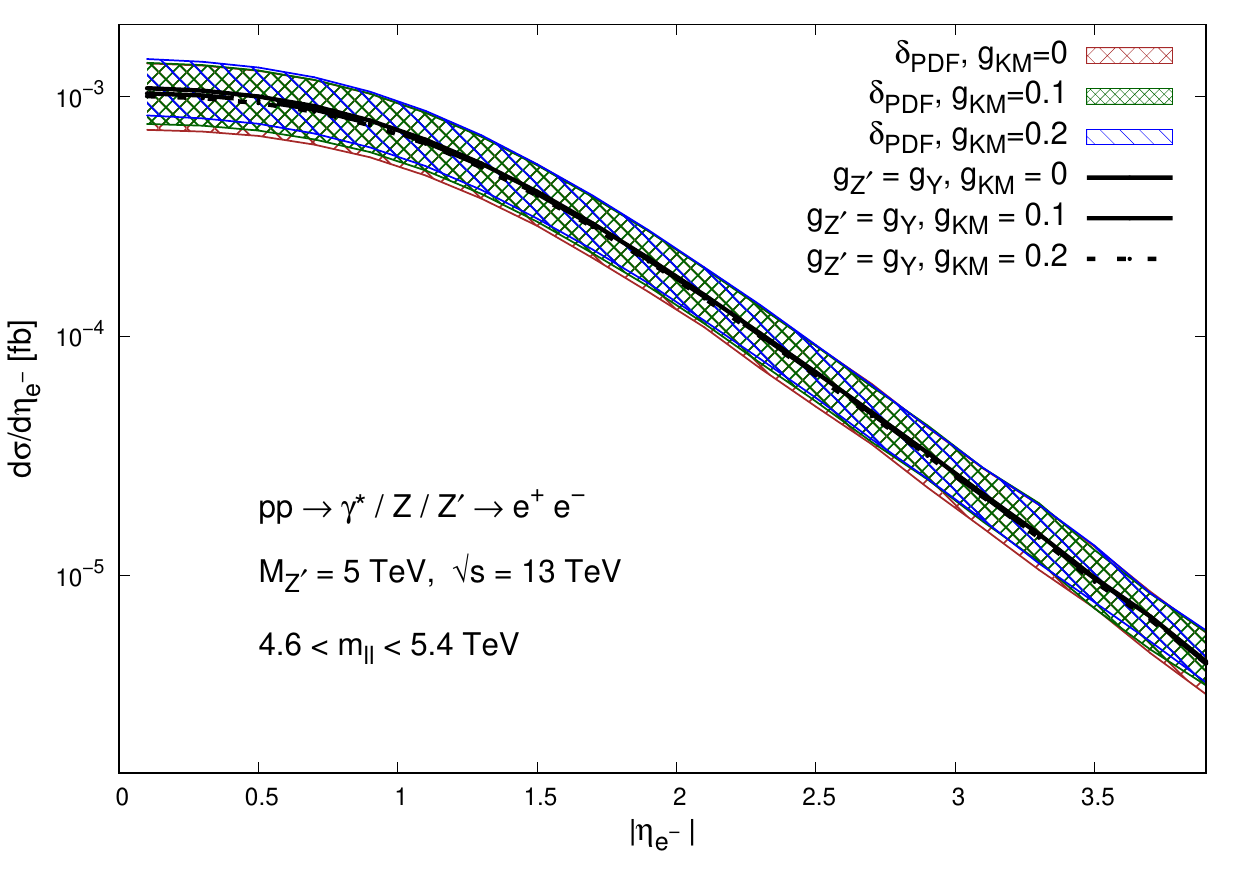}
\caption{Left: Differential cross-section as a function of $\cos \theta$, where the $\theta$ angle is defined in the Collins-Soper frame. Right: Pseudorapidity distribution $\eta_e$ of the outgoing electron. Error bands with different hatching represent the CT18NNLO PDF uncertainties induced on the cross section, evaluated at the 90\% CL. Central predictions are represented by lines with different dashing.}
\label{fig:angular}
\end{figure}

Finally, in Fig.~\ref{fig:angular} we show the impact of varying the $g_{Z'}$ and $g_{KM}$ parameters on the $\cos{\theta}$ distribution, and on the pseudorapidity $\eta_e$ spectrum of the final-state electron. The angle $\theta$ is defined in the Collins-Soper frame~\cite{Collins:1977iv}. While there is almost no impact on the $\eta_e$ spectrum, a distortion of the central value of the $\cos\theta$ angular distribution is observed when $g_{KM}$ is progressively increased from 0 to 0.2. As we shall see in the next section, this is reflected by the $A_{FB}$ distribution. However, all these effects in the $\cos{\theta}$ distribution are buried by the almost complete overlap of the PDF uncertainty bands for $g_{KM}=0,0.1, 0.2$.         

\subsection{Forward-Backward Asymmetry $A_{FB}$ distribution results \label{DY-AFB}}
\begin{figure}
\includegraphics[width=0.55\textwidth]{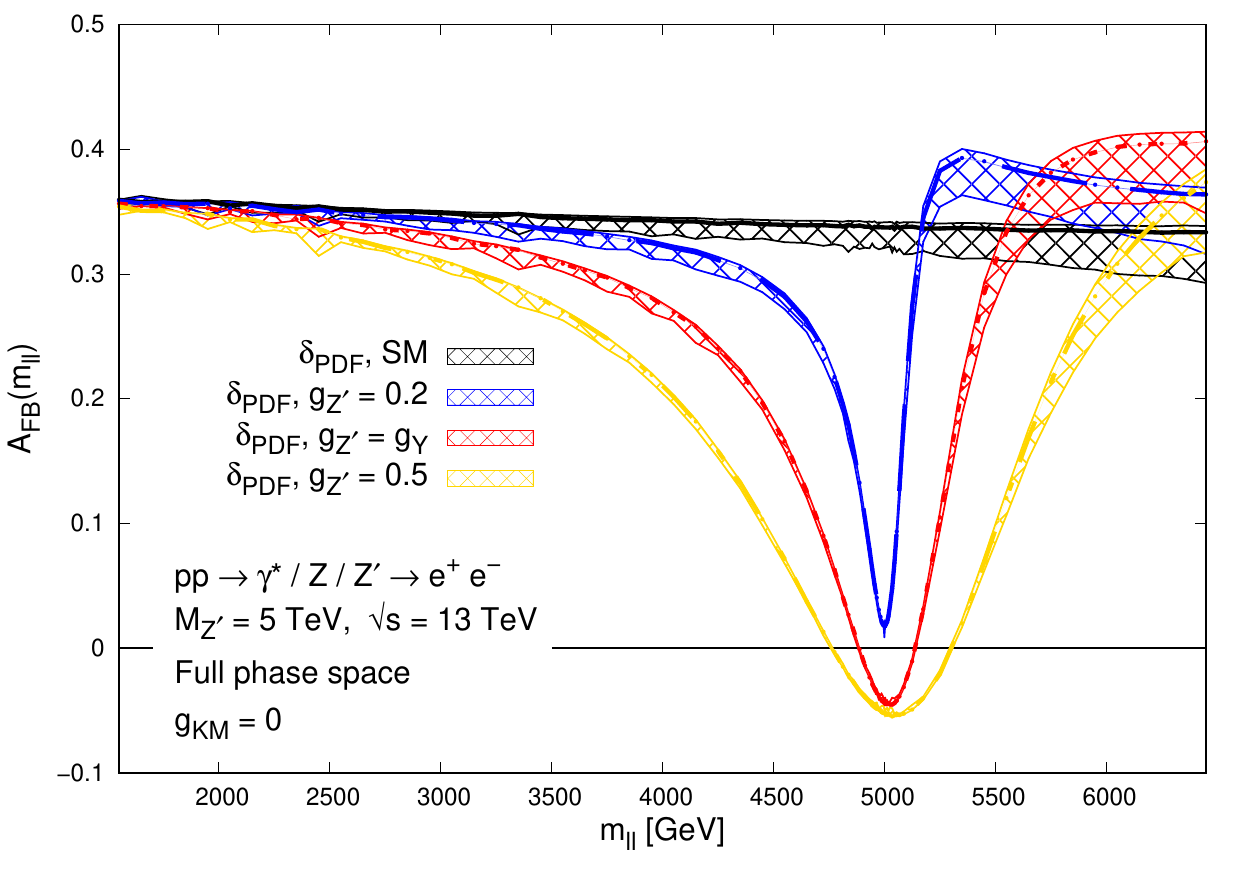}
\includegraphics[width=0.55\textwidth]{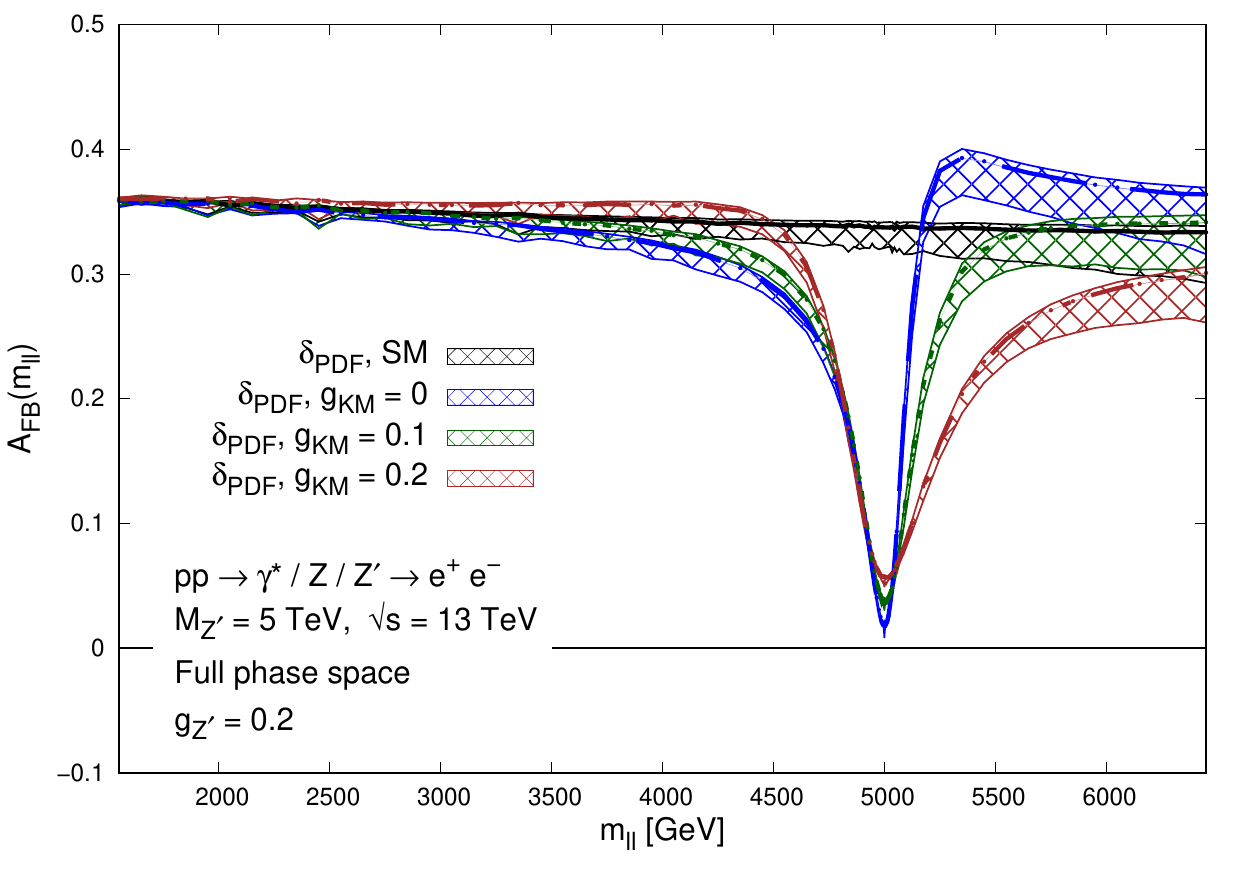}
\includegraphics[width=0.55\textwidth]{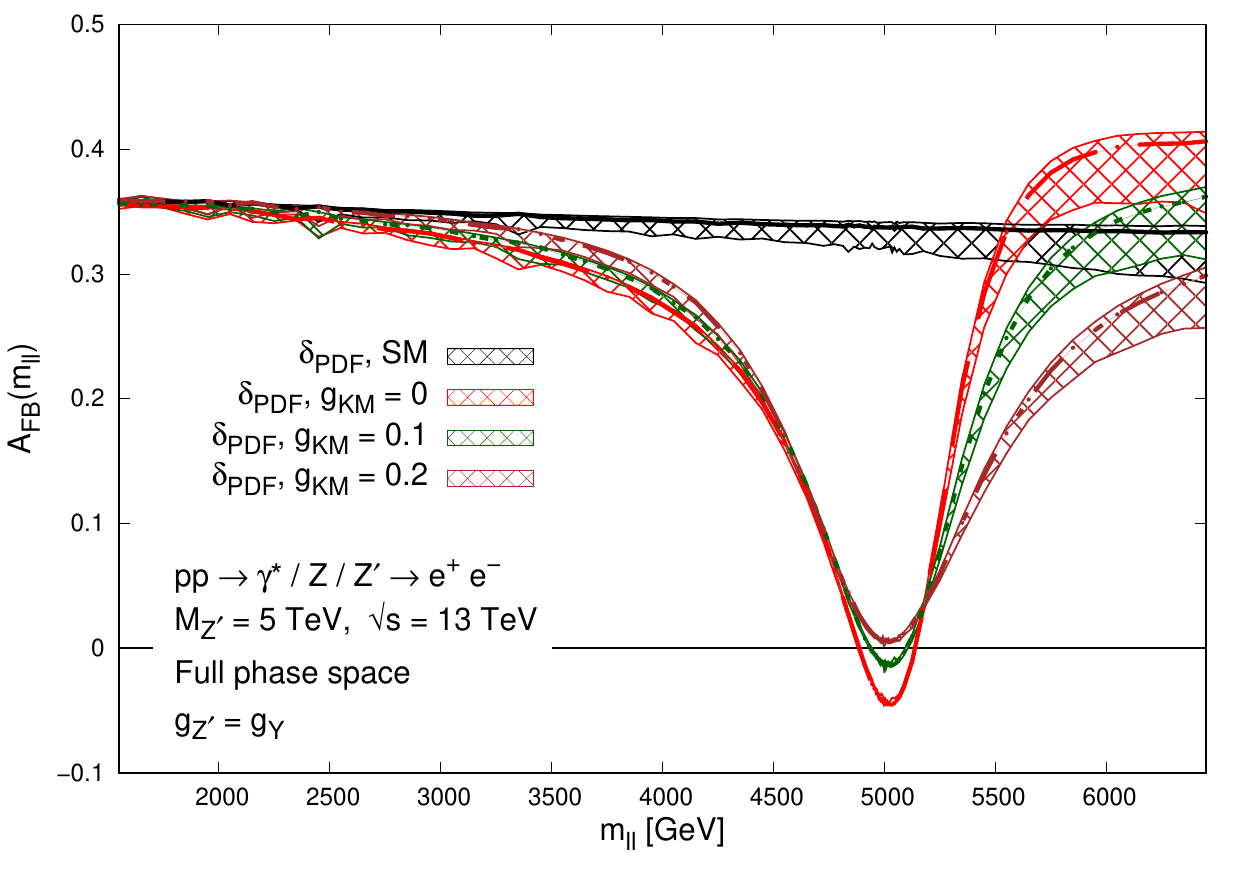}
\caption{Forward-backward asymmetry ($A_{FB}$) shown as a function of the dilepton invariant mass ($m_{ll}$) for different values of $g_{Z'}$ and $g_{KM}$. The black hatched band represents the Standard Model prediction. The solid black line at $A_{FB} = 0$ is introduced for reference. The CT18NNLO PDF uncertainties at the 90\% CL are represented by hatched bands. Central predictions are shown as lines with different dashing.}
\label{fig:afb-5TeV}
\end{figure}

As anticipated in the previous sections, a more suitable variable to explore the $Z'$ parameters in DY is the forward-backward asymmetry $A_{FB}$ distribution. 
This observable is a function of the chiral quark and lepton couplings to the mediating gauge bosons and is very sensitive to $Z'$ properties. $A_{FB}$ is expressed in terms of the difference between the forward and backward angular contributions to the cross section, normalized to the total cross section. The forward and backward 
directions are represented by the angle $\theta$ between the negatively charged final-state lepton and the initial-state quark in the dilepton center-of-mass frame. $A_{FB}$ is defined as 
\begin{equation}
    A_{FB} = \frac{d\sigma_F - d\sigma_B}{d\sigma_F + d\sigma_B}
\end{equation}
where the forward ($d\sigma_F$) and backward ($d\sigma_B$) contributions are constructed by integrating the differential cross section over the forward and backward halves of the angular phase space 
\begin{equation}
    d\sigma_F = \int_0^1 \frac{d\sigma}{d\cos{\theta_l}} \,d\cos{\theta_l}, \;\;\; d\sigma_B = \int_{-1}^{0} \frac{d\sigma}{d\cos{\theta_l}} \,d\cos{\theta_l}\,.
\end{equation}
$\theta_l$ is the angle between the final-state lepton and initial-state quark in the Collins-Soper frame. In our analysis, we study the $A_{FB}$ distribution as a function of $m_{ll}$. In Fig.~\ref{fig:afb-5TeV} we show the $A_{FB}$ over the same invariant mass range adopted before. The presence of a $Z'$ would lead to a distortion in the $A_{FB}$ which becomes wider as $g_{Z'}$ increases. We calculated both PDF and scale uncertainties for the $A_{FB}$ distribution. Scale uncertainties are essentially invisible in Fig.~\ref{fig:afb-5TeV}, while PDF uncertainties are attenuated around the peak as compared to the distributions studied in the previous section, but are still large at high $m_{ll}$.
Fig.~\ref{fig:afb-5TeV} shows the sensitivity of $A_{FB}$ to simultaneous variations of the $g_{Z'}$ and $g_{KM}$ parameters. The distortions induced on the asymmetry tend to become shallower and wider as the kinetic mixing increases. Moving away from the resonance, when $g_{KM}\neq 0$, the discriminatory power of $A_{FB}$ appears to be attenuated by PDF uncertainties.  

\section{Conclusions\label{conclusions}}
In this work, we studied properties of heterotic string derived $Z'$s in the TeV scale and explored their dynamics at the LHC. As a case study, we selected a $Z'$ of mass $M_{Z'}=5$ TeV.
We analyzed the impact of a gauge kinetic-mixing term in the Lagrangian and its interplay with the $Z'$ coupling. We explored the $Z'$ parameter space the model in presence of kinetic mixing and performed a thorough phenomenological analysis in which we used precision calculations at NNLO in QCD to predict kinematic distributions for the $Z'$ in the Drell-Yan process at the LHC. In particular, we exploited the sensitivity of forward-backward asymmetry $A_{FB}$ distributions in Drell-Yan to further explore the parameter space and the interplay between the $Z'$ coupling and the kinetic-mixing parameter.   
Moreover, we estimated theory uncertainties on the distributions from perturbative (scale dependence) and nonperturbative (PDFs in the proton) sources in QCD. Proton PDFs still remain one of the major sources of uncertainty in $Z'$ searches. This is ascribed to the fact that the kinematic domain of heavy gauge boson production is sensitive, and is impacted by, PDFs at large $x$ where uncertainties are still large due to lack of robust constraints from experimental data~\cite{Brady:2011hb,Amoroso:2022eow}.  
The motivation to consider the particular $U(1)_{Z'}$ combination in Eq.~\ref{uzpwuzeta} stems from its extraction from a string derived heterotic--string model~\cite{Faraggi:2014ica}, in which it is the unique combination that may remain unbroken down to low scales. Its preservation as an unbroken gauge symmetry down to the TeV scale emanates
from the role that it can play in suppressing some dangerous operators, like proton decay mediating operators, and furthermore by the role it can play in generating the electroweak symmetry breaking itself. Further analysis in that direction mandates the development of the interpolation tools between the string and electroweak scales and the extraction of further data from the string models. We note that while in this paper we treated the kinetic-mixing term as a free parameter, its computation, as that of many of the related parameters, can be obtained directly from the string models~\cite{Dienes:1996zr}, hence increasing their predictive power.
In this paper, we studied the case of $Z'$ vector bosons that have an $E_6$ embedding. The advantage of this choice is that it facilitates gauge unification at the unification scale~\cite{Faraggi:2013nia}. 
Such extra string inspired $Z'$ models have been discussed extensively in the literature since the mid--eighties~\cite{Zwirner:1987kxa,Hewett:1988xc,Leike:1998wr,King:2020ldn}, 
but the combination given in Eq.~\ref{uzpwuzeta} is obtained in the string derived model of ref.~\cite{Faraggi:2014ica}. Naturally, the string constructions can give rise to $U(1)$ symmetries with different 
characteristics and these have been of some interest in the literature~\cite{Pati:1996fn,Faraggi:2000cm,Coriano:2007ba,Faraggi:2011xu}. The range of possibilities 
is more model dependent. The extra $Z'$s in these cases can arise from flavour dependent $U(1)$ 
symmetries, as well as $U(1)$ symmetries that arise from the hidden sector of the heterotic--string. 
Naturally, the limits that we discussed in this paper will not apply to those cases (see for instance the discussion relative to $Z'$ boson searches in the PDG~\cite{ParticleDataGroup:2022pth} and references therein), as they do not couple 
universally to all the three families. On the other hand, there may exist other constraints in these cases
arising from flavour non--universality and the need to break flavour dependent symmetries to produce 
viable fermion masses. Nevertheless, these cases do represent interesting alternatives to the 
family universal $U(1)$ and we will return to them in future work. 



\subsection*{Acknowledgments}
The work of MG and AM is supported by the National Science foundation under Grant no. 2112025. The work of AF is supported in part by 
a Weston visiting professorship at the Weizmann Institute of Science. AF would like to thank Doron Gepner for discussions and the Department
of Particle Physics and Astrophysics for hospitality.

\appendix

\section{Gauge boson rotation matrix components \label{appendix:rot}} 

In this section we report the exact components of the matrix $\mathcal O_{gauge}$ which rotates the neutral gauge field basis to the physical basis as in Eq. \ref{eq::rotmatrix}.
\begin{equation} \tag{A1}
\begin{split}
    &\mathcal O_{11} = g_Y / g, \quad \mathcal O_{21} = g_2 / g, \quad \mathcal O_{31} = 0, \\
    &\mathcal O_{12} = -g_2 \frac{v^2 \left (f_1 + \sqrt{f_1^2 + 4 g^2 x_z^2} \right) - 2 x_z^2}{\sqrt{g^2 \left[ v^2 \left( f_1 + \sqrt{f_1^2 + 4 g^2 x_z^2} \right) - 2 x_z^2 \right]^2 + x_z^2 \left( f_2 - \sqrt{f_1^2 + 4 g^2 x_z^2} \right)^2}}, \\
    &\mathcal O_{22} = g_Y \frac{v^2 \left (f_1 + \sqrt{f_1^2 + 4 g^2 x_z^2} \right) - 2 x_z^2}{\sqrt{g^2 \left[ v^2 \left( f_1 + \sqrt{f_1^2 + 4 g^2 x_z^2} \right) - 2 x_z^2 \right]^2 + x_z^2 \left( f_2 - \sqrt{f_1^2 + 4 g^2 x_z^2} \right)^2}}, \\
    &\mathcal O_{32} = \frac{x_z \left( f_2 - \sqrt{f_1^2 + 4 g^2 x_z^2} \right)}{\sqrt{g^2 \left[ v^2 \left( f_1 + \sqrt{f_1^2 + 4 g^2 x_z^2} \right) - 2 x_z^2 \right]^2 + x_z^2 \left( f_2 - \sqrt{f_1^2 + 4 g^2 x_z^2} \right)^2}}, \\
    &\mathcal O_{13} = -g_2 \frac{v^2 \left( f_1 - \sqrt{f_1^2 + 4 g^2 x_z^2} \right) - 2 x_z^2}{\sqrt{g^2 \left[ v^2 \left( f_1 - \sqrt{f_1^2 + 4 g^2 x_z^2} \right) - 2 x_z^2 \right]^2 + x_z^2 \left( f_2 + \sqrt{f_1^2 + 4 g^2 x_z^2} \right)^2}}, \\
    &\mathcal O_{23} = g_Y \frac{v^2 \left( f_1 - \sqrt{f_1^2 + 4 g^2 x_z^2} \right) - 2 x_z^2}{\sqrt{g^2 \left[ v^2 \left( f_1 - \sqrt{f_1^2 + 4 g^2 x_z^2} \right) - 2 x_z^2 \right]^2 + x_z^2 \left( f_2 + \sqrt{f_1^2 + 4 g^2 x_z^2} \right)^2}}, \\
    &\mathcal O_{33} = \frac{x_z \left( f_2 + \sqrt{f_1^2 + 4 g^2 x_z^2} \right)}{\sqrt{g^2 \left[ v^2 \left( f_1 - \sqrt{f_1^2 + 4 g^2 x_z^2} \right) - 2 x_z^2 \right]^2 + x_z^2 \left( f_2 + \sqrt{f_1^2 + 4 g^2 x_z^2} \right)^2}}
\end{split}
\end{equation}
where $f_1 = N_z - g^2v^2$ and $f_2 = N_z + g^2v^2$. To include the effects of kinetic mixing on $\mathcal O_{gauge}$, we make the substitutions $x_z \rightarrow x_z'$ and $N_z \rightarrow N_z'$ as defined in Eq. \ref{eq:xz_km}.

\section{$Z' \rightarrow H^+H^-$ decay rate\label{appendix:Zp-rates}}

The decay rate of the $Z'$ into charged Higgs bosons $Z' \rightarrow H^+H^-$ in presence of kinetic mixing is given by 
\begin{equation}
    \Gamma_{Z' \rightarrow H^+H^-} = \frac{g^2_{Z'H^+H^-}}{16\pi^2M_{Z'}^5} \left[ M_{Z'}^2 (M_{Z'} - 2M_{H^\pm}) (2M_{H^\pm} + M_{Z'}) \right]^{3/2},
\end{equation}
where the structure of the coupling is
\begin{equation}
\begin{split}
    g_{Z'H^+H^-} = &\left( \frac{v_1}{v} \right)^2 (g_{Z'} z_{H_1} + \varepsilon g_2 \cos \theta_W) + \left( \frac{v_2}{v} \right)^2 (-g_{Z'} z_{H_2} + \varepsilon g_2 \cos \theta_W) \\ &+ (g_{KM} + \varepsilon g_Y \sin \theta_W) \left( \left( \frac{v_1}{v} \right)^2 Y_{H_1} - \left( \frac{v_2}{v} \right)^2 Y_{H_2} \right).
    \end{split}
\end{equation}

\bibliographystyle{JHEP}

\providecommand{\href}[2]{#2}\begingroup\raggedright\endgroup

\end{document}